\documentclass[aps, 10pt, prd, notitlepage, twocolumn, superscriptaddress, nofootinbib, floatfix]{revtex4-1}
\usepackage[pdftex]{graphicx}              
\usepackage{amsmath}
\usepackage{amssymb}
\usepackage{amsfonts}
\usepackage[utf8]{inputenc}
\usepackage[T1]{fontenc}
\usepackage{mathrsfs}
\usepackage{anyfontsize}
\usepackage{enumerate}
\usepackage{tensor}
\usepackage{soul}

\usepackage[linktocpage,breaklinks]{hyperref}
\usepackage[usenames,dvipsnames]{xcolor}

\usepackage{tensor}
\usepackage{bm}

\usepackage{graphicx}
\graphicspath{{./figures/}}

\usepackage{natbib}
\usepackage{hyperref}
\usepackage{orcidlink}
\usepackage{verbatim}

\hypersetup{colorlinks=true,
            citecolor=RoyalBlue,
            linkcolor=RoyalBlue,
            urlcolor=RoyalBlue}

\newcommand{\be}{\begin{eqnarray}}
\newcommand{\ee}{\end{eqnarray}}

\newcommand{\ICASU}{\affiliation{Illinois Center for Advanced Studies of the Universe, University of Illinois at Urbana-Champaign, Urbana, Illinois 61801, USA}}

\newcommand{\Physics}{\affiliation{Department of Physics, University of Illinois at Urbana-Champaign, 1110 West Green St, Urbana, IL 61801, USA}}

\newcommand{\Konrad}{\affiliation{Programa de Matem\'atica, Fundaci\'on Universitaria Konrad Lorenz, 110231 Bogot\'a, Colombia.}}

\newcommand{\Princeton}{\affiliation{Department of Physics, Princeton University, Princeton, NJ, 08544, USA}}

\newcommand{\Astro}{\affiliation{Department of Astronomy, University of Illinois at Urbana-Champaign, 1002 West Green Street, Urbana, IL 61801, USA}}

\newcommand{\NCSA}{\affiliation{National Center for Supercomputing Applications, 1205 W Clark St, Urbana, IL 61801, USA}}

\begin{document}

\title{Blandford--Znajek Process in Quadratic Gravity}

\author{Jameson Dong\orcidlink{0000-0002-6081-0238}}
\Physics

\author{Nicol\'as Pati\~no\orcidlink{0000-0002-3155-9750}}
\Physics

\author{Yiqi Xie\orcidlink{0000-0002-8172-577X}}
\Physics
\ICASU

\author{\\ Alejandro~C\'ardenas-Avenda\~no\orcidlink{0000-0001-9528-1826}}
\ICASU
\Konrad
\Princeton

\author{Charles~F.~Gammie\orcidlink{0000-0001-7451-8935}~}
\Physics
\ICASU
\Astro
\NCSA

\author{Nicol\'as~Yunes\orcidlink{0000-0001-6147-1736}~}
\Physics
\ICASU

\date{\today}


\begin{abstract}
The Blandford--Znajek process, which uses a magnetized plasma to extract energy from a rotating black hole, is one of the leading candidates for powering relativistic jets. In this work, we investigate the Blandford--Znajek process in two well-motivated quadratic gravity theories: scalar Gauss--Bonnet and dynamical Chern--Simons gravity.  We solve analytically for a split-monopole magnetosphere to first order in the small-coupling approximation and second relative order in the slow-rotation approximation. The extracted power at fixed spin and magnetic flux is enhanced in scalar Gauss--Bonnet and reduced in dynamical Chern--Simons gravity, compared to general relativity. We find that there is a degeneracy between spin and the coupling constants of the theories at leading order in the slow rotation approximation that is broken at higher orders. 

\end{abstract}


\maketitle

\section{Introduction}
\label{sec:intro}

\allowdisplaybreaks[4]

Direct electromagnetic extraction of the rotational energy of supermassive black holes (BH) via the Blandford-Znajek (BZ) process~\cite{1977MNRAS.179..433B} is a plausible power source for relativistic jets in many active galactic nuclei (AGN)~\cite{Narayan:2011eb,Steiner:2012ap,Blandford:2018iot,Chen:2021ffu}. In the BZ process, the ergosphere of a rotating BH is threaded by a poloidal magnetic field embedded in a highly conducting plasma. As the magnetic field lines are frame-dragged, a toroidal field forms, and the work done by the BH on the field lines leads to the extraction of its rotational energy. This theoretical framework for relativistic jets is supported by modeling of Event Horizon Telescope (EHT) observations~\cite{EventHorizonTelescope:2019pgp,EventHorizonTelescope:2021srq,EventHorizonTelescope:2020dlu}. Comparison of models with EHT observations of M87* favors those models in which M87's jet originates in a low-density, magnetically-dominated region (the ``funnel'') over the poles of the black hole. 

Over the past three decades, the BZ process has been extensively studied in general relativity (GR). The analytical studies (e.g.~\cite{1977MNRAS.179..433B,Beskin:2000qe,McKinney_2004,Tanabe:2008wm,Tchekhovskoy_2010,Pan:2015haa,Grignani:2018ntq,Armas:2020mio}) compute the fields perturbatively, and the associated energy flux is therefore found, under certain assumptions, to a particular order in the BH's spin. For example,~\cite{Armas:2020mio} recently calculated the field configuration to third relative order\footnote{In this work, the term ``relative order in spin'' refers to the scaling with spin, relative to the leading-order expression in a slow-rotation expansion. For example, the third relative order field configuration includes the poloidal magnetic field at the third order (which is the leading-order contribution), plus both the toroidal magnetic field at the fourth order, and the BZ power at the fifth order.} in the spin parameter using matched asymptotic expansions and abandoning the assumption that the field variables are smooth in the BH's spin. GR magnetohydrodynamic (GRMHD) simulations have also shown that, for slowly rotating BHs, the structure of the time-averaged funnel magnetic field matches the analytic solution of Blandford and Znajek~\cite{Komissarov:2001sjq,Komissarov:2004ms,McKinney_2004,Tchekhovskoy_2010,Penna:2013rga}. Simulations also enable the study of rapidly rotating black holes, but there are no analytical models to compare these results, and they are computationally expensive~\cite{Talbot:2020zkb}.  

Since the BZ process depends on astrophysics (through the magnetosphere) and the theory of gravity (through the exterior BH spacetime metric)~\cite{Komissarov:2004ms,Ruiz:2012te,Toma:2014kva}, studying the process and its observational consequences may probe gravity in the strong-field regime. In modified gravity, however, the BZ mechanism has been much less studied than in GR, and when considered, it has been studied only analytically. For instance, in~\cite{Bambi:2012zg,Pei:2016kka,Konoplya:2021qll,Banerjee:2020ubc}, the BZ power was computed to leading order in the BH's spin, either for an agnostic, parametrically deformed (``bumpy''~\cite{Collins:2004ex}) BH metric~\cite{Bambi:2012zg,Pei:2016kka,Konoplya:2021qll} or for a theory-specific (Kerr-Sen) BH metric~\cite{Banerjee:2020ubc}.

Moreover, previous studies in modified gravity have all considered a magnetosphere in which the rotation frequency of the electromagnetic (EM) field maximizes the power output of the BZ process and followed the procedure devised in~\cite{Tchekhovskoy_2010} for the Kerr metric. Using a magnetosphere that maximizes the BZ power is a good approximation for the magnetosphere dynamics around Kerr BHs~\cite{Beskin:2000qe}. However, it is unknown whether that assumption applies generically to other spacetimes, and a careful investigation of magnetospheric structure in non-Kerr BH spacetimes is needed. 
In this paper, we address these difficulties by studying in great detail the BZ process in two quadratic gravity theories: scalar Gauss--Bonnet (sGB) gravity~\cite{Kanti:1995vq,Yunes:2011we,Ayzenberg:2014aka,Maselli:2015tta} and dynamical Chern--Simons (dCS) gravity~\cite{Alexander:2009tp,Yagi:2012ya,Maselli:2017kic}. Both theories are well-motivated extensions of GR from the effective theory standpoint~\cite{Yunes:2011we} and arise in low-energy expansions of quantum gravity theories~\cite{Boulware:1985wk,Kanti:1995vq,Alexander:2004xd,Taveras:2008yf,Weinberg:2008hq,Alexander:2009tp}. 

We solve the governing equations of the magnetosphere around BHs described by these quadratic theories analytically (to first order in the small-coupling approximation and second relative order in the small-rotation approximation) by combining the solution strategies presented in~\cite{1977MNRAS.179..433B,McKinney_2004,Armas:2020mio} for a split-monopole configuration. Our results suggest that using a magnetosphere that maximizes the BZ power remains a good approximation for the magnetosphere dynamics around modified gravity. We also find that the power of energy extraction from the BH, compared to the predictions of GR, is enhanced in sGB gravity and quenched in dCS gravity.  At leading order, we find that there is a degeneracy between the BH's parameters, namely the spin and the parameter (coupling constant) that controls the modification from GR. This degeneracy makes it difficult to use the BZ mechanism to place limits on the coupling parameters of the theory, even in the presence of high-quality data. We then show that this degeneracy is broken at higher orders in the perturbative scheme. 

This paper is organized as follows:
Section~\ref{sec:bz} reviews the mathematical formulation of the BZ process. 
Section~\ref{sec:gr_bz} presents the solution to the BZ process in GR following a simplified strategy based on~\cite{1977MNRAS.179..433B,McKinney_2004,Armas:2020mio} and discusses its advantages and limitations. 
Section~\ref{sec:qg} reviews the BH solutions in sGB and dCS gravity, solves the BZ process in these theories, compares the result with the prediction of GR, and ends with a detailed explanation of the differences found. Section~\ref{sec:implications} describes in detail a degeneracy between the BH parameters that appears at leading order and discuss its implications to future studies of the BZ process in modified theories of gravity. Section~\ref{sec:discussion} summarizes and discusses future work. Throughout the paper, we use geometric units with $G_{\rm N}=1=c$ and the metric signature $(-,+,+,+)$.

\section{The Blandford--Znajek Process}
\label{sec:bz}

The BZ process assumes a stationary, axisymmetric magnetosphere -- composed of an electromagnetic field and a highly conducting plasma -- around a rotating BH~\cite{1977MNRAS.179..433B}. The EM field energy is large compared to the plasma rest-mass density everywhere except close to the equatorial plane, where matter accretes in a high-density disk. We assume the split-monopole configuration, where the disk is considered as a thin current sheet, and the magnetic field lines are considered to be asymptotically radial as they cross two-spheres far from the BH. Off the disk, the dominance of the EM field implies the force-free condition (e.g., see~\cite{10.1093/mnras/167.3.457}):
\begin{align}
    F_{\mu\nu}J^\nu=0, \label{eq:force_free}
\end{align}
where $F_{\mu\nu}$ is the Faraday tensor of the EM field, and $J^\nu=\nabla_\mu F^{\mu\nu}$ is the four-current. 
The disk appears as a discontinuity to the EM field, and the magnetic field lines cross the equatorial plane only through the central BH. This split-monopole configuration in GR has been extensively used for the analytical study of the BZ process~\cite{1977MNRAS.179..433B,McKinney_2004,Tanabe:2008wm,Tchekhovskoy_2010,Pan:2015haa,Grignani:2018ntq,Armas:2020mio}, and its analytic solution has been shown to agree with numerical simulations~\cite{Komissarov:2001sjq,Komissarov:2004ms,McKinney_2004}.
In the following, we adopt Gralla and Jacobson's notation~\cite{Gralla:2014yja} as it can be easily applied to theories beyond GR. 

In Boyer--Lindquist (BL) coordinates $(t,r,\theta,\phi)$, a stationary and axisymmetric metric can be decomposed in the following form
\begin{align}
    ds^2=g_{\mu\nu}dx^\mu dx^\nu=g^T_{AB}dx^Adx^B+g^P_{ab}dx^adx^b, \label{eq:metric_22decomp}
\end{align}
where $g^T_{AB}$ is referred to as the ``toroidal metric,'' and $g^P_{ab}$ is referred to as the ``poloidal metric;'' in other words, the toroidal coordinates $(t,\phi)$ are indexed with uppercase letters, and the poloidal coordinates $(r,\theta)$ with lowercase letters. This particular decomposition is not unique to GR and can be performed for rotating BHs in several theories of gravity~\cite{Xie:2021bur}, including the quadratic gravity theories of interest here. Gralla and Jacobson~\cite{Gralla:2014yja} showed that a stationary, axisymmetric, and force-free EM field can always be represented by
\begin{align}
    F_{tr}=&-F_{rt}=\Omega\,\partial_r\psi, \label{eq:faraday_tr}\\
    F_{t\theta}=&-F_{\theta t}=\Omega\,\partial_\theta\psi, \label{eq:faraday_ttheta}\\
    F_{r\phi}=&-F_{\phi r}=\partial_r\psi, 
    \label{eq:faraday_rphi}\\
    F_{\theta\phi}=&-F_{\phi\theta}=\partial_\theta\psi, \label{eq:faraday_thetaphi}\\
    F_{r\theta}=&-F_{\theta r}=\frac{I}{2\pi}\sqrt{\frac{g^P}{-g^T}}, \label{eq:faraday_rtheta}
\end{align}
with all other components zero. Here $\psi$, $I$, and $\Omega$ are functions of $(r,\theta)$, and $g^T$ and $g^P$ are the determinants of the toroidal metric and the poloidal metric, respectively. The quantities $2\pi\psi$ and $I$ measure the magnetic flux and the electric current through a surface bounded by the loop of revolution at $(r,\theta)$, respectively, and $\Omega$ (which is constant along field lines) measures the rotation frequency of magnetic field lines being dragged by the rotation of the BH. We refer to $\psi$ as the ``poloidal flux function,'' $I$ as the ``poloidal current function,'' and $\Omega$ as the ``rotation frequency.''
A similar description (e.g., see~\cite{1977MNRAS.179..433B,McKinney_2004,Tanabe:2008wm,Pan:2015haa}) can be made in terms of a toroidal vector potential $A_\phi$ and a toroidal magnetic field $B_T$ or $B^\phi$, instead of the flux function $\psi$ and the current function $I$, respectively. These descriptions are related by $d\psi=dA_\phi$ and $I=2\pi B_T=-2\pi g^T B^\phi$. 

Inserting Eqs.~\eqref{eq:faraday_tr}--\eqref{eq:faraday_rtheta} into the force-free condition of Eq.~\eqref{eq:force_free}, the $t$ and $\phi$ components become
\begin{align}
    \partial_rI\,\partial_\theta\psi=&\partial_\theta I\,\partial_r\psi, \label{eq:ipsi}\\
    \partial_r\Omega\,\partial_\theta\psi=&\partial_\theta\Omega\,\partial_r\psi, \label{eq:omegapsi}
\end{align}
which may also be interpreted as $I$ and $\Omega$ being functions of $\psi$. On the other hand, the $r$ and $\theta$ components of Eq.~\eqref{eq:force_free} can be combined into the stream equation~\cite{Gralla:2014yja}:
\begin{align}
    \nabla_\mu(|\eta|^2\nabla^\mu\psi)+\Omega'(\eta\cdot dt)|\nabla\psi|^2-\frac{II'}{4\pi^2g^T}=0, \label{eq:stream}
\end{align}
where the prime denotes a $\psi$ derivative, and $\eta\equiv d\phi-\Omega\,dt$.
Due to parity, finding a solution in the northern hemisphere ($0<\theta<\pi/2$) would be sufficient, as $\psi$, $I$, and $\Omega$ in the southern hemisphere mirror the northern hemisphere solution. 

The total EM energy flux extracted from the BH, also known as the BZ power, is~\cite{1977MNRAS.179..433B,Gralla:2014yja}
\begin{align}
    P=&-\int I\Omega\,d\psi\notag\\=&4\pi \int_0^{\pi/2}\left[\Omega\,(\Omega_\mathrm{H}-\Omega)\,(\partial_\theta\psi)^2\sqrt{\frac{g_{\phi\phi}}{g_{\theta\theta}}}\,\right]\bigg|_{r=r_\mathrm{H}}\,d\theta, \label{eq:power}
\end{align}
where $r_\mathrm{H}$ is the horizon radius which can be found as the outermost solution to $g^T=0$, and 
\begin{align}
\Omega_\mathrm{H}\equiv-\left.\frac{g_{t\phi}}{g_{\phi\phi}} \right|_{r=r_\mathrm{H}}
\end{align}
is the horizon angular frequency. 
The functional form of Eq.~\eqref{eq:power} indicates that the energy flux is directed outward on the horizon when $0<\Omega<\Omega_\mathrm{H}$, and it is sometimes assumed that the field rotation frequency equals half of the horizon angular frequency, i.e.~$\Omega=\Omega_\mathrm{H}/2$ (for instance, see~\cite{Tchekhovskoy_2010,Konoplya:2020hyk}). We will not make that assumption in this work. The importance of not making this assumption will become explicit when we study the BZ mechanism in sGB and dCS gravity in Sec.~\ref{sec:qg}.

Prescribing the boundary conditions for Eqs.~\eqref{eq:ipsi}--\eqref{eq:stream} turns out to be a delicate job. Here we adopt the boundary conditions from a recent work by Armas et al.~\cite{Armas:2020mio}:
\begin{align}
    \psi=0,&\quad\theta=0, \label{eq:flux_pole}\\
    \psi=\psi_0,&\quad\theta=\pi/2, \label{eq:flux_equator}\\
    \psi~\textrm{finite},&\quad r=r_\mathrm{H}, \label{eq:flux_horizon}\\
    I=2\pi(\Omega-\Omega_\mathrm{H})\,\partial_\theta\psi\sqrt{\frac{g_{\phi\phi}}{g_{\theta\theta}}},&\quad r=r_\mathrm{H}, \label{eq:znajek_horizon}\\
    I=-2\pi\Omega\,\partial_\theta\psi\,\sin\theta,&\quad r\rightarrow\infty , \label{eq:znajek_infinity} \\
    \psi~\textrm{finite},&\quad r\rightarrow\infty, 
    \label{eq:flux_infinity}
\end{align}
where $\psi_0$ is a constant. The condition stipulated by Eq.~\eqref{eq:flux_pole} is required by the physical interpretation of $\psi$: at the north pole, the surface over which the magnetic flux is measured shrinks to a zero size, so the flux function there should be set to zero. The condition~\eqref{eq:flux_equator} is a restatement of the split-monopole assumption that no magnetic field line crosses the disk, and therefore $\psi_0$ determines the magnetic flux through the horizon. Equations~\eqref{eq:flux_horizon} and \eqref{eq:znajek_horizon} come from the requirement that the EM field strength, $F_{\mu\nu}$, be finite when measured by a timelike observer traveling across the horizon.

Equation~\eqref{eq:znajek_horizon} is the Znajek condition~\cite{1977MNRAS.179..457Z}, which is equivalent to requiring a finite toroidal magnetic field, $B^\phi$, in horizon-penetrating coordinates~\cite{McKinney_2004}. Gralla and Jacobson~\cite{Gralla:2014yja} have extended this condition so that it holds as long as the horizon is a Killing horizon generated by $\partial_t+\Omega_\mathrm{H}\partial_\phi$. The Znajek condition mapped to null future infinity becomes Eq.~\eqref{eq:znajek_infinity}~\cite{Penna:2015qta,Armas:2020mio}. There is no need to adapt Eq.~\eqref{eq:znajek_infinity} to a generic metric since we are considering spacetimes that are asymptotically flat. The conditions given by Eqs.~\eqref{eq:znajek_horizon} and \eqref{eq:znajek_infinity} can be derived, up to a sign, by directly solving Eq.~\eqref{eq:stream} on the horizon and at the infinity with the assumption that $\psi$, $I$, and $\Omega$ are all finite there. The sign is fixed by assuming that the energy flow is outwardly directed on the horizon and at the infinity~\cite{MacDonald:1982zz,Gralla:2014yja,Gralla:2015vta}. Finally, Eq.~\eqref{eq:flux_infinity} matches the field at infinity with Michel's flat-space monopole solution~\cite{1973ApJ...180L.133M}.

Together with these boundary conditions, Eqs.~\eqref{eq:ipsi}--\eqref{eq:stream}, first derived by Blandford and Znajek~\cite{1977MNRAS.179..433B}, are therefore all one needs to solve for the fields (either analytically or numerically). However, the only known \emph{exact} solution to these equations is a generalization of Michel's monopole solution~\cite{1973ApJ...180L.133M} in the Schwarzschild spacetime~\cite{1977MNRAS.179..433B}, which lacks astrophysical interest as no energy can be extracted. Therefore, perturbation methods are typically applied to study this process analytically. 

\section{The Blandford--Znajek Process in General Relativity}
\label{sec:gr_bz}

In this section, we revisit the BZ process in GR and present a simplified self-contained rederivation of the known solutions~\cite{1977MNRAS.179..433B,Armas:2020mio}. We start by writing the Kerr metric in BL coordinates:
\begin{align}
    ds^2=&-\left(1-\frac{2Mr}{\Sigma}\right)dt^2 + \frac{\Sigma}{\Delta}\,dr^2 + \Sigma\,d\theta^2 \notag\\
    &+ \frac{1}{\Sigma}\left[(r^2+a^2)^2-a^2\Delta\sin^2\theta\right]\sin^2\theta\,d\phi^2 \notag\\
    &- \frac{4Mar\sin^2\theta}{\Sigma}\,dt\,d\phi, \label{eq:lineelem_kerrbl}
\end{align}
where $\Delta\equiv r^2-2Mr+a^2$, $\Sigma\equiv r^2+a^2\cos^2\theta$, and $a$ and $M$ denote the BH's spin and mass, respectively. The event horizon is located at
\begin{align}
    r_\mathrm{H}=M+\sqrt{M^2-a^2}, \label{eq:rh_gr}
\end{align}
where the angular frequency is
\begin{align}
    \Omega_\mathrm{H}=\frac{a}{2Mr_\mathrm{H}}. \label{eq:omegah_gr}
\end{align}
The ergosphere is located at
\begin{align}
    r_{\mathrm{ergo}}=M+\sqrt{M^2-a^2\cos^2\theta}. \label{eq:rergo_gr}
\end{align}
Let us now consider a slowly-rotating Kerr BH with dimensionless spin parameter $\chi\equiv a/M\ll1$, and expand the field variables in powers of $\chi$. Let us assume that the field variables are smooth functions of $\chi$ at $\chi=0$, and the following functional form for the expansions
\begin{align}
    \psi=&\psi^{(0)}(x,\theta)+\chi^2\psi^{(2)}(x,\theta)+\mathcal{O}(\chi^4), \label{eq:expand_gr_psi}\\
    I=&\chi I^{(1)}(x,\theta)+\chi^3 I^{(3)}(x,\theta)+\mathcal{O}(\chi^5),  \label{eq:expand_gr_i}\\
    \Omega=&\chi \Omega^{(1)}(x,\theta)+\chi^3 \Omega^{(3)}(x,\theta)+\mathcal{O}(\chi^5), \label{eq:expand_gr_omega}
\end{align}
where we have introduced $x\equiv r/M$ as a dimensionless radial coordinate.

Following~\cite{Armas:2020mio}, let us now define what we mean by ``relative order in spin'' formally. When expanding in small spins, some functions will have some $\chi$ dependence to leading order. A term of $N$th relative spin order then means a term that is $\chi^N$ smaller than the leading-order term. With this in mind then, $\psi^{(0)}$, $I^{(1)}$, and $\Omega^{(1)}$ are zeroth relative order (leading order); $\psi^{(1)}$, $I^{(2)}$, and $\Omega^{(2)}$ are first relative order, and the terms shown in Eqs.~\eqref{eq:expand_gr_psi}--\eqref{eq:expand_gr_omega} are the field expansion up to second relative order. 

At leading order, the stream Eq.~\eqref{eq:stream} reads
\begin{align}
    L\psi^{(0)}=0,\label{eq:stream0_gr}
\end{align}
where $L$ is a separable differential operator defined by~\cite{Petterson:1974bt}
\begin{align}
    L = \frac{\partial}{\partial x}\left[\left(1-\frac{2}{x}\right)\frac{\partial}{\partial x}\right] + \frac{\sin\theta}{x^2}\frac{\partial}{\partial\theta}\left(\frac{1}{\sin\theta}\frac{\partial}{\partial\theta}\right). \label{eq:lop_mid}
\end{align}
Imposing the boundary conditions of Eqs.~\eqref{eq:flux_pole}--\eqref{eq:flux_infinity}, one obtains
\begin{align}
    \psi^{(0)} = \psi_0(1-\cos\theta), \label{eq:psi0_gr}
\end{align}
which is the exact monopole solution in the northern hemisphere. Note that here we imposed the horizon condition of Eq.~\eqref{eq:flux_horizon} at $x=2$, instead of at $x=r_\mathrm{H}/M$. Clearly, this will not affect the solution at $\mathcal{O}(\chi^0)$ because the difference between $x = 2$ and $x=r_\mathrm{H}/M$ is of $\mathcal{O}(\chi^2)$; such correction will be accounted for when we study the solution at higher order in $\chi$. In the following, we will always impose the horizon conditions, i.e.~Eqs.~\eqref{eq:flux_horizon} and \eqref{eq:znajek_horizon}, at $x=2$ instead of $x=r_\mathrm{H}/M$.
By inserting Eq.~\eqref{eq:psi0_gr} into Eqs.~\eqref{eq:ipsi} and \eqref{eq:omegapsi}, one finds that $I^{(1)}$ and $\Omega^{(1)}$ depend solely on $\theta$ and can be determined by the Znajek conditions of Eqs.~\eqref{eq:znajek_horizon} and \eqref{eq:znajek_infinity} to obtain
\begin{align}
    I^{(1)}=-2\pi\psi_0\Omega^{(1)}\sin^2\theta, \quad \Omega^{(1)}=\frac{1}{8M}. \label{eq:iomega0_gr}
\end{align}
Note that when $x\rightarrow\infty$, Eqs.~\eqref{eq:psi0_gr}--\eqref{eq:iomega0_gr} match Michel's flat-space solution~\cite{1973ApJ...180L.133M}.

Given that $\Omega_\mathrm{H}=\chi/(4M)+\mathcal{O}(\chi^3)$ for Kerr BHs, by comparing it to Eq.~\eqref{eq:iomega0_gr}, one finds
\begin{align}
    \Omega=\frac{1}{2}\Omega_\mathrm{H}+\mathcal{O}(\chi^3). \label{eq:omega0_half}
\end{align}
In fact, $\Omega\approx\Omega_\mathrm{H}/2$ is a common feature of the BZ process around a slowly-rotating Kerr BH~\cite{Thorne:1986iy,Beskin:2000qe,Komissarov:2004ms,McKinney_2004,Tchekhovskoy_2010}.
As a consequence, Tchekhovskoy et al.~\cite{Tchekhovskoy_2010} suggested to take $\Omega=\Omega_\mathrm{H}/2$ as the solution to the field rotation frequency at leading order, and referred to it as the ``energy argument,'' given that the BZ power is maximized by this rotation frequency at leading order. Recent studies of the BZ process in modified gravity theories (e.g.~\cite{Pei:2016kka,Konoplya:2020hyk,Banerjee:2020ubc}) adopted this suggestion and gave estimates of the BZ power in the slow-rotation limit without solving for the fields. This treatment, however, is not justified since $\Omega\approx\Omega_\mathrm{H}/2$ may not hold in general, i.e., for other theories of gravity. In addition, if higher orders in the spin parameter are considered, the approximation displayed in Eq.~\eqref{eq:omega0_half} is insufficient, as we will show later.  

We now go to next order in the perturbative scheme. Similar to the leading order treatment, the second relative order stream Eq.~\eqref{eq:stream} takes the form 
\begin{align}
    L\psi^{(2)}=-\psi_0 \frac{x+2}{x^4}\cos\theta\sin^2\theta,\label{eq:stream2_gr}
\end{align}
and by requiring that $\psi$ satisfy Eqs.~\eqref{eq:flux_pole}--\eqref{eq:flux_infinity}, the solution is simply
\begin{align}
    \psi^{(2)}=\psi_0\,f(x)\cos\theta\sin^2\theta, \label{eq:psi2_gr}
\end{align}
where~\cite{McKinney_2004}
\begin{align}
    f(x)=&\frac{1}{8}x^2(2x-3)\left[\mathrm{Li}_2\left(\frac{2}{x}\right)+\ln\left(\frac{2}{x}\right)\ln\left(1-\frac{2}{x}\right)\right]\notag\\
    &+ \frac{1}{12}(6x^2-3x-1)\ln\left(\frac{2}{x}\right)-\frac{1}{6}x^2(x-1)\notag\\&+\frac{11}{72}+\frac{1}{3x}, \label{eq:fr_gr}
\end{align}
and $\mathrm{Li}_2(x)\equiv-\int_0^1(1/t)\log(1-xt)\,dt$ is the second polylogarithm function. At the boundaries,
\begin{align}
    f(2)=\frac{-49+6\pi^2}{72},\quad
    f(x)\Big|_{x\rightarrow\infty}\sim\frac{1}{4x}. \label{eq:fr_limit_gr}
\end{align}
Solving Eqs.~\eqref{eq:ipsi}--\eqref{eq:omegapsi} with the conditions given by Eqs.~\eqref{eq:znajek_horizon}--\eqref{eq:znajek_infinity}, we find
\begin{align}
    I^{(3)}=&-2\pi\psi_0\sin^2\theta\left[\Omega^{(3)}+\frac{\cos^2\theta}{4M}f(x)\right],\\ \Omega^{(3)}=&\frac{1}{32M}\left(1+\frac{67-6\pi^2}{36}\sin^2\theta\right). \label{eq:iomega2_gr}
\end{align}
As mentioned above, at this order $\Omega$ deviates from $\Omega_{\mathrm{H}}/2$. Therefore, the assumption that the rotation frequency takes the value which maximizes the power is not true at higher order in spin. To second relative order, the BZ power, Eq.~\eqref{eq:power}, is
\begin{align}
    P=\frac{\pi}{24}\frac{\psi_0^2}{M^2}\chi^2+\frac{\pi(56-3\pi^2)}{1080}\frac{\psi_0^2}{M^2}\chi^4+\mathcal{O}(\chi^6), \label{eq:power_gr}
\end{align}
which agrees with the results first presented in~\cite{Tanabe:2008wm}. Blandford and Znajek~\cite{1977MNRAS.179..433B} first solved the field variables up to $\psi^{(2)}$, $I^{(1)}$, and $\Omega^{(1)}$, and they evaluated the BZ power to leading order. The next-to-leading-order BZ power was obtained by Tanabe and Nagataki~\cite{Tanabe:2008wm} without the solutions for $I^{(3)}$ and $\Omega^{(3)}$, which were recently found by Armas et al.~\cite{Armas:2020mio}. 

The above perturbative procedure, however, cannot be extended to higher orders. In particular, the $\mathcal{O}(\chi^4)$ flux function $\psi^{(4)}$ will not satisfy the boundary condition in Eq.~\eqref{eq:flux_infinity}, and instead, it will diverge at large $r$~\cite{Tanabe:2008wm}. In addition, terms of the form $\mathcal{O}(|\chi|^3)$ and $\mathcal{O}(\chi^4\log|\chi|)$ will appear in the expansion of $\psi$~\cite{Armas:2020mio}, meaning that the field will not be a smooth function of $\chi$ any longer. We will now comment on these two issues. 

A consistent treatment to the slowly-rotating, split-monopole BZ process was first attempted by Grignani et al.~\cite{Grignani:2018ntq} and was recently resolved by Armas et al.~\cite{Armas:2020mio} using three distinct slow-rotation expansions connected by matched asymptotics. These expansions are referred to as ``near,'' ``mid,'' and ``far'' with respect to distance between the horizon and where the expansion applies. These regions also represent the three spatial regimes separated by the inner and outer light surfaces~\cite{Komissarov:2004ms,Nathanail:2014aua}.
Under this scheme, the derivation we presented earlier should be thought of as the solution of the mid expansion, except that our boundary conditions for the horizon and infinity should be imposed at the near expansion and the far expansion, respectively. As a consequence, the mid expansion no longer requires a finite field at the boundaries, and the divergence of $\psi^{(4)}$ at large $r$ can now be buffered by some well-behaved term in the far expansion.

Regarding the problem of the smoothness of the fields, Armas et al. suggested to consider all powers of $\chi$ in the first place and check whether other forms of dependence (e.g.,  $\chi^4\log|\chi|$) should be included every time the solution at an order is found~\cite{Armas:2020mio}. In GR, up to the second relative order, Armas et al. found that the smoothness assumption holds true, and the near and far solutions can be extended from the mid solution by taking $r=r_\mathrm{H}$ and $r\rightarrow\infty$, respectively. 
In the next section, we will show that this argument holds in quadratic gravity theories as well, such that the derivation presented above holds, and it is not necessary to use the procedure presented by Armas et al.~\cite{Armas:2020mio}. 

We have shown that including the boundary condition Eq.~\eqref{eq:znajek_infinity} correctly solves for the fields. Consider, for example, the step we took from Eq.~\eqref{eq:psi0_gr} to Eq.~\eqref{eq:iomega0_gr}, where the conditions given by Eqs.~\eqref{eq:znajek_horizon} and~\eqref{eq:znajek_infinity} were used to determine $I^{(1)}$ and $\Omega^{(1)}$. If Eq.~\eqref{eq:znajek_infinity} was not provided, then one could only determine $I^{(1)}$ as a function of $\Omega^{(1)}$ (or the opposite, i.e.,~$\Omega^{(1)}$ as a function of $I^{(1)}$). When going to next order in the stream equation [Eq.~\eqref{eq:stream2_gr}], one finds that the source term would also be a function of $\Omega^{(1)}$. In general, this new Eq.~\eqref{eq:stream2_gr} would no longer be compatible with the boundary condition of Eq.~\eqref{eq:flux_infinity} unless some constraint was put on the source term. Once this required constraint was found, one could combine it with the requirement that $\Omega$ and $I$ be finite to solve $\Omega^{(1)}$ and $I^{(1)}$, and eventually $\psi^{(2)}$. The solution to higher orders would be similar, with the feature of needing to determine $\Omega^{(n)}$ and $I^{(n)}$ with the next order stream equation, as presented in~\cite{McKinney_2004,Tanabe:2008wm,Pan:2015haa,Grignani:2018ntq} for example. 
Under such a scheme, since we work to the second relative order, solving for $I^{(3)}$ and $\Omega^{(3)}$ would require the problematic $\psi^{(4)}$.

In the seminal derivation of this process~\cite{1977MNRAS.179..433B}, Blandford and Znajek used Eq.~\eqref{eq:znajek_infinity} as a shortcut to match Michel's solution~\cite{1973ApJ...180L.133M} at leading order in the spin expansion, without necessarily implying that it would hold at all orders. However, in subsequent works (e.g.~\cite{McKinney_2004,Tanabe:2008wm,Pan:2015haa,Grignani:2018ntq}) the condition that $\Omega$ and $I$ are finite at infinity was used instead of Eq.~\eqref{eq:znajek_infinity}. Following~\cite{Armas:2020mio} we adopted Eq.~\eqref{eq:znajek_infinity}, as this boundary condition is equivalent to requiring finite field variables and no incoming energy from the infinity (i.e., an isolated magnetosphere).

In quadratic gravity theories, not using the condition of Eq.~\eqref{eq:znajek_infinity} would lead to the same issues that appear in GR. As we will explain later in Sec.~\ref{sec:qg_bz2} and with great detail in Appendix~\ref{apd:test}, for the quadratic theories considered in this work, we cannot apply the scheme proposed by Armas et al.~\cite{Armas:2020mio}. Therefore, we choose to use the condition of Eq.~\eqref{eq:znajek_infinity} to avoid these problems and keep our derivations simple. 

\section{The Blandford--Znajek Process in Quadratic Gravity}
\label{sec:qg}

\subsection{Rotating Black Holes in Quadratic Gravity}

Perhaps the most well-studied cases of theories that correct GR through higher curvature terms are scalar Gauss Bonnet (sGB) gravity~\cite{Campbell:1991kz} and dCS gravity~\cite{Jackiw:2003pm,Alexander:2009tp}. Quadratic gravity theories result as extensions of GR from the effective field theory standpoint~\cite{Yunes:2011we} and arise in the low-energy expansions of quantum gravity theories, in which scalar fields and higher-order curvature terms appear as corrections to GR~\cite{Boulware:1985wk,Kanti:1995vq,Alexander:2004xd,Taveras:2008yf,Weinberg:2008hq,Alexander:2009tp}. 

In sGB and dCS gravity, a dynamical massless scalar $\vartheta_\mathrm{sGB}$ and pseudo-scalar $\vartheta_\mathrm{dCS}$, respectively, are coupled to the gravitational field through quadratic-in-curvature scalar invariants. These theories are defined in vacuum by adding to the Einstein-Hilbert action a scalar field coupled to the metric as follows~\cite{Yagi:2015oca}:
\begin{align}
    S_{\mathrm{sGB}}=&\int d^4x\sqrt{-g}\,\bigg[-\frac{1}{2}(\nabla_\mu\vartheta_\mathrm{sGB})(\nabla^\mu\vartheta_\mathrm{sGB})\notag\\
    &+\alpha_\mathrm{sGB}\vartheta_\mathrm{sGB}(R^2-4R_{\mu\nu}R^{\mu\nu}+R_{\mu\nu\rho\sigma}R^{\mu\nu\rho\sigma})\bigg],\\
    S_{\mathrm{dCS}}=&\int d^4x\sqrt{-g}\,\bigg[-\frac{1}{2}(\nabla_\mu\vartheta_{{\mathrm{dCS}}})(\nabla^\mu\vartheta_{{\mathrm{dCS}}})\notag\\
    &-\frac{\alpha_\mathrm{dCS}}{4}\vartheta_\mathrm{dCS}\,^*\!R_{\mu\nu\rho\sigma}R^{\mu\nu\rho\sigma}\bigg],
\end{align}
where the quadratic scalar invariants $R^2$, $R_{\mu\nu}R^{\mu\nu}$, the Kretschmann scalar $R_{\mu\nu\rho\sigma}R^{\mu\nu\rho\sigma}$, and the Pontryagin density ${}^*R_{\mu\nu\rho\sigma}R^{\mu\nu\rho\sigma}$, where ${}^*R_{\mu\nu\rho\sigma}=\frac{1}{2}\tensor{\epsilon}{^{\alpha\beta}_{\rho\sigma}}R_{\mu\nu\alpha\beta}$ is the dual of the Riemann, are coupled through the coupling constants $\alpha_\mathrm{sGB}$ and $\alpha_\mathrm{dCS}$, respectively. The most stringent constraints to date from gravitational-wave observations are (to 90\% confidence): $\alpha_\mathrm{sGB}^{1/2}\leq5.6\,\mathrm{km}$~\cite{Nair:2019iur} and $\alpha_\mathrm{dCS}^{1/2}\leq8.5\,\mathrm{km}$~\cite{Silva:2020acr}.

In sGB, the scalar field is coupled to a quadratic curvature invariant, which is parity even, and therefore, the spherical solutions in this theory are different from Schwarzschild. On the other hand, in dCS, the curvature invariant is parity odd, and therefore, any spherically symmetric solution in GR is also a solution in dCS gravity, e.g.~the Schwarzschild solution~\cite{Yunes:2009hc}. 

Currently, exact closed-form solutions that represent rotating BHs in sGB and dCS gravity do not exist. Therefore, in this work, we use the small-coupling and slow-rotation approximate solutions found in sGB~\cite{Yunes:2011we,Pani:2011gy,Ayzenberg:2014aka,Maselli:2015tta} and in dCS gravity~\cite{Yunes:2009hc,Pani:2011gy,Yagi:2012ya,Maselli:2017kic}. The small-coupling approximation treats the metric solutions in both theories as deformed from the Kerr solution by deviations proportional to the dimensionless coupling parameter
\begin{align}
    \zeta_q\equiv\frac{\alpha_q^2}{\kappa M^4}\ll1, 
\end{align}
where $q\in\{\mathrm{sGB,dCS}\}$ refers to either theory, $\kappa=(16 \pi)^{-1}$, and $M$ is the mass of the compact object. We will use the approximate solutions up to $\mathcal{O}(\zeta_q,\chi^5)$, which are presented in Appendix \ref{apd:metric} for completeness\footnote{These solutions are different from those in~\cite{Maselli:2017kic} because that paper used Hartle--Thorne coordinates, and we use BL coordinates.}. We note that both solutions in BL coordinates follow the decomposition presented in Eq.~\eqref{eq:metric_22decomp}. 

Now, let us summarize some of the BH characteristics in these solutions that we will use later, up to $\mathcal{O}(\zeta_{q},\chi^3)$. First, the horizon radial locations are
\begin{align}
    r_{\mathrm{H},\mathrm{sGB}}=&r_{\mathrm{H},\mathrm{GR}}-\frac{49}{40}\zeta_\mathrm{sGB} M-\frac{277}{960}\zeta_\mathrm{sGB}\chi^2 M,\label{eq:rh_gb}\\
    r_{\mathrm{H},\mathrm{dCS}}=&r_{\mathrm{H},\mathrm{GR}}-\frac{915}{28672}\zeta_\mathrm{dCS}\chi^2 M,\label{eq:rh_cs}
\end{align}
where $r_{\mathrm{H},\mathrm{GR}}$ is the horizon radius of the Kerr metric as given in Eq.~\eqref{eq:rh_gr}. 
As in GR, the horizons are generated by the Killing vector $\partial_t+\Omega_\mathrm{H}\partial_\phi$, where the horizon angular frequencies are
\begin{align}
    \Omega_{\mathrm{H},\mathrm{sGB}}=&\Omega_{\mathrm{H},\mathrm{GR}} + \frac{\zeta_\mathrm{sGB}\chi}{M}\bigg(\frac{21}{80} -\frac{21103}{201600}\chi^2\bigg), \label{eq:omegah_gb}\\
    \Omega_{\mathrm{H},\mathrm{dCS}}=&\Omega_{\mathrm{H},\mathrm{GR}} - \frac{\zeta_\mathrm{sGB}\chi}{M}\bigg(\frac{709}{28672} +\frac{169}{24576}\chi^2\bigg), \label{eq:omegah_cs}
\end{align}
where $\Omega_{\mathrm{H},\mathrm{GR}}$ is the horizon angular frequency of the Kerr metric as given in Eq.~\eqref{eq:omegah_gr}. 
Finally, the ergospheres are also modified, with radii now given by
\begin{align}
    r_{\mathrm{ergo},\mathrm{sGB}}=&r_{\mathrm{ergo},\mathrm{GR}}-\frac{49}{40}\zeta_\mathrm{sGB} M\notag\\&+\frac{277}{960}\zeta_\mathrm{sGB}\chi^2 M \left(1-\frac{850}{277}\sin^2\theta\right),\\
    r_{\mathrm{ergo},\mathrm{dCS}}=&r_{\mathrm{ergo},\mathrm{GR}}\notag\\&-\frac{915}{28672}\zeta_\mathrm{dCS}\chi^2 M \left(1+\frac{2836}{915}\sin^2\theta\right),
\end{align}
where $r_{\mathrm{ergo},\mathrm{GR}}$ is the ergosphere radius of the Kerr metric as given in Eq.~\eqref{eq:rergo_gr}. 
Using the modified location of the horizon and the ergosphere, we develop a resummed version of the approximated metric solutions that recovers the exact Kerr solution as $\zeta_{q}\rightarrow0$ and shifts the coordinate singularity to the respective value of the horizon. The details are presented in Appendix \ref{apd:resum}. 

\subsection{The Blandford--Znajek Process in Quadratic Gravity to Leading Order in Spin}
\label{sec:qg_bz0}

Let us consider the BZ process around BHs in sGB and dCS gravity. 
As in GR, we solve the force-free conditions in Eqs.~\eqref{eq:ipsi}--\eqref{eq:omegapsi} constrained by the boundary conditions of Eqs~\eqref{eq:flux_pole}--\eqref{eq:flux_infinity} and evaluate the BZ power using Eq.~\eqref{eq:power}.
The field expansions are now
\begin{align}
    \psi_q=&\psi_q^{(0,0)}+\chi^2\psi_q^{(0,2)}\notag\\
    &+\zeta_q\psi_q^{(1,0)}+\zeta_q\chi^2\psi_q^{(1,2)}+\mathcal{O}(\zeta_q^2,\chi^4),\\
    I_q=&\chi I_q^{(0,1)}+\chi^3 I_q^{(0,3)} \notag\\
    &+\zeta_q\chi I_q^{(1,1)}+\zeta_q\chi^3 I_q^{(1,3)}+\mathcal{O}(\zeta_q^2,\chi^5),\\
    \Omega_q=&\chi \Omega_q^{(0,1)}+\chi^3\Omega_q^{(0,3)} \notag\\
    &+\zeta_q\chi\Omega_q^{(1,1)}+\zeta_q\chi^3\Omega_q^{(1,3)}+\mathcal{O}(\zeta_q^2,\chi^5),
\end{align}
where the integer pair $(m,n)$ stands for the $m$th order in each coupling constant $\zeta_q$ and the $n$th order in the spin $\chi$. As GR is recovered when these couplings vanish, $\psi_q^{(0,n)}$, $I_q^{(0,n)}$, and $\Omega_q^{(0,n)}$ are the same as $\psi^{(n)}$, $I^{(n)}$, and $\Omega^{(n)}$ in Sec.~\ref{sec:gr_bz}. 
Thus, we only need to solve for $\psi_q^{(1,n)}$, $I_q^{(1,n)}$, and $\Omega_q^{(1,n)}$ for each theory.

Let us first consider the solutions at leading order in spin. The stream Eq.~\eqref{eq:stream} reads
\begin{align}
    L\psi_q^{(1,0)}=0, \label{eq:stream0_qg}
\end{align}
and by imposing the boundary conditions of Eqs.~\eqref{eq:flux_pole}--\eqref{eq:flux_horizon} and \eqref{eq:flux_infinity}, the solution is
\begin{align}
    \psi_q^{(1,0)}=0. \label{eq:psi0_qg}
\end{align}
Note that although Eq.~\eqref{eq:stream0_qg} is the same as the leading order GR stream equation in Eq.~\eqref{eq:stream0_gr}, the resulting solution is different. This is because the GR solution $\psi_q^{(0,0)}$ has already accounted for all the monopole charge $\psi_0$, so the charge condition Eq.~\eqref{eq:flux_equator} cancels any further corrections taking the same form of $\psi_q^{(0,0)}$.
From Eqs.~\eqref{eq:ipsi} and \eqref{eq:omegapsi}, together with the conditions in Eqs.~\eqref{eq:znajek_horizon} and \eqref{eq:znajek_infinity}, we obtain
\begin{align}
    I^{(1,1)}_\mathrm{sGB}=-2\pi\psi_0\Omega^{(1,1)}_\mathrm{sGB}\sin^2\theta,&\quad \Omega^{(1,1)}_\mathrm{sGB}=\frac{21}{160M}, \label{eq:omega0_gb}\\
    I^{(1,1)}_\mathrm{dCS}=-2\pi\psi_0\Omega^{(1,1)}_\mathrm{dCS}\sin^2\theta,&\quad \Omega^{(1,1)}_\mathrm{dCS}=-\frac{709}{57344M}. \label{eq:omega0_cs}
\end{align}
The corrections to the BZ power, according to Eq.~\eqref{eq:power}, are therefore
\begin{align}
    P^{(1,2)}_\mathrm{sGB}=&\frac{7\pi}{80}\frac{\psi_0^2}{M^2}, \label{eq:power0_gb}\\
    P^{(1,2)}_\mathrm{dCS}=&-\frac{709\pi}{86016}\frac{\psi_0^2}{M^2}. \label{eq:power0_cs}
\end{align}

Combining Eqs.~\eqref{eq:iomega0_gr}, \eqref{eq:omega0_gb}--\eqref{eq:omega0_cs}, and \eqref{eq:omegah_gb}--\eqref{eq:omegah_cs}, we find that
\begin{align}
    \Omega_q=\frac{1}{2}\Omega_{\mathrm{H},q}+\mathcal{O}(\zeta_q^2,\chi^3).
\end{align}
This result is analogous to Eq.~\eqref{eq:omega0_half} but extended to quadratic gravity, and it indicates that the field rotation frequency takes the value that maximizes the BZ power at leading order in spin, as in the GR case to the same order. From Eqs.~\eqref{eq:power0_gb}--\eqref{eq:power0_cs}, together with Eq.~\eqref{eq:power_gr} and \eqref{eq:omegah_gb}--\eqref{eq:omegah_cs}, the BZ power can then be written as
\begin{align}
    P_{q}=\frac{\pi}{6}\psi_0^2\Omega_{\mathrm{H},q}^2+\mathcal{O}(\zeta_q^2,\chi^4). \label{eq:power0_half}
\end{align}
This expression coincides with the result presented in~\cite{Konoplya:2021qll} for the maximal BZ power when using generic parametrized BH metrics to $\mathcal{O}(\Omega_\mathrm{H}^2)$. The above derivation provides a proof and shows that the value of the rotation frequency does not have to be assumed, as it is a consequence of the magnetosphere dynamics.
Furthermore, as mentioned in~\cite{Konoplya:2021qll}, from Eq.~\eqref{eq:power0_half} and the corrections to the Kerr horizon angular frequency, i.e., Eqs.~\eqref{eq:omegah_gb}--\eqref{eq:omegah_cs}, one can phenomenologically infer the main contributions from the metric coefficients to the BZ power.

The derivation shown above suggests that Eq.~\eqref{eq:omega0_half} should hold as a consequence of the magnetosphere dynamics in all modified theories of gravity that admit BH solutions that can be described as continuous deformations of the Schwarzschild metric. Generically, at leading order in spin the stream Eq.~\eqref{eq:stream} should take the form of Eq.~\eqref{eq:stream0_gr}:
\begin{align}
    L_\mathrm{mod} \psi^{(0)}_\mathrm{mod}=0\,,
    \label{eq:gen-arg-eq}
\end{align}
where the subscript $\mathrm{mod}$ stands for ``modified theory,'' and the superscript $(n)$ stands for a term of ${\cal{O}}(\chi^n)$, following the notation introduced in Sec.~\ref{sec:gr_bz}. Both $L_\mathrm{mod}$ and $\psi^{(0)}_\mathrm{mod}$ contain a GR part and a non-GR part that depends on the coupling constants of the modified theory. Regardless of the details of the modified theory, $L_\mathrm{mod}$ is of ${\cal{O}}(\chi^0)$, so the metric that one uses to calculate it must be spherically symmetric.
In BL coordinates, such a metric is diagonal, and its angular sector is just the metric of the two-sphere, i.e.,
\begin{align}
    g_{\theta\theta,\mathrm{mod}}^{(0)}=r^2,\quad
    g_{\phi\phi,\mathrm{mod}}^{(0)}=r^2\sin^2\theta. \label{eq:metric_angular}
\end{align}	
Therefore, Eq.~\eqref{eq:gen-arg-eq} should still be separable, and its angular sector should still be the same as that of $L$ in Eq.~\eqref{eq:lop_mid}. As a result, the leading order in spin stream equation should still accept the solution
\begin{align}
    \psi^{(0)}_\mathrm{mod}=\psi_0(1-\cos\theta)\,.
\end{align}
As shown above, solving Eqs.~\eqref{eq:ipsi} and \eqref{eq:omegapsi}, together with the conditions in Eqs.~\eqref{eq:znajek_horizon} and \eqref{eq:znajek_infinity}, and inserting the angular metric components in Eq.~\eqref{eq:metric_angular}, one obtains
\begin{align}
    \Omega_\mathrm{mod}^{(1)}=\frac{1}{2}\Omega_\mathrm{H,mod}.
\end{align}
Thus,
\begin{align}
    \Omega_\mathrm{mod}=\frac{1}{2}\Omega_\mathrm{H,mod}+\mathcal{O}(\chi^3),
\end{align}
for a generic theory of gravity that describes continuous deformations of the Schwarzschild metric. The argument presented above, however, is not a proof because a rigorous statement would require that we understand the behavior of the metric in the near horizon and the far field, or alternatively that we can develop a resummation of the metric and show that this behavior is unimportant. Without specifying a particular modified theory of gravity, it is not clear how to establish those results, but this, in any case, is outside the scope of this paper.

According to Eqs.~\eqref{eq:power0_gb}--\eqref{eq:power0_cs} and \eqref{eq:power_gr}, given a BH of fixed mass and spin, the relative corrections to the BZ power, with respect to GR, by sGB and dCS are
\begin{align}
    \frac{P_\mathrm{sGB}-P_\mathrm{GR}}{P_\mathrm{GR}}\approx& \, 2\zeta_\mathrm{sGB}, \label{eq:prel_gb}\\
    \frac{P_\mathrm{dCS}-P_\mathrm{GR}}{P_\mathrm{GR}}\approx& -0.2\zeta_\mathrm{dCS}. \label{eq:prel_cs}
\end{align}
Thus, the correction is one order of magnitude larger in sGB than in dCS gravity. In addition, there is a sign difference so that the power is enhanced in sGB gravity and quenched in dCS gravity, with respect to the prediction of GR. 

The difference in the corrections found, both in magnitude and in sign, can be traced back to the different corrections to the BH metric in the vicinity of the horizon. 
At leading order, the toroidal metric components of BHs in both theories can be written as 
\begin{align}
    g_{tt,q}=&1-2M/r + \zeta_q k_q(r/M) + \mathcal{O}(\zeta_q^2,\chi^2),\label{eq:gtt_qg}\\
    g_{\phi\phi,q}=&r^2\sin^2\theta + \mathcal{O}(\zeta_q^2,\chi^2),\\
    g_{t\phi,q}=&-\chi\big[2M/r + \zeta_q l_q(r/M)\big]\sin^2\theta + \mathcal{O}(\zeta_q^2,\chi^3),\label{eq:gtphi_qg}
\end{align}
where $k_q(r)$ and $l_q(r)$ are different functions for sGB and dCS that can be obtained by comparing Eqs.~\eqref{eq:gtt_qg}--\eqref{eq:gtphi_qg} with the BH solutions provided in Appendix~\ref{apd:metric}. 
Given that $\Omega_\mathrm{H}\equiv-g_{t\phi}/g_{\phi\phi}|_{r=r_\mathrm{H}}$, and $r_\mathrm{H}$ is the solution to $g^T=0$, we find
\begin{align}
    r_{\mathrm{H},q}=&2M\big[1-\zeta_q k_q(2)\big] + \mathcal{O}(\zeta^2,\chi^2),\\
    \Omega_{\mathrm{H},q}=&\frac{\chi}{4M}\left[1+3\zeta_q k_q(2)+\zeta_q l_q(2)\right] + \mathcal{O}(\zeta_q^2,\chi^3).
\end{align}
Then using Eq.~\eqref{eq:power0_half}, we can write 
\begin{align}
    \frac{P_q-P_\mathrm{GR}}{P_\mathrm{GR}}=2\zeta_q [3k_q(2)+l_q(2)]+\mathcal{O}(\zeta_q^2,\chi^2). \label{eq:prel_qg}
\end{align}
To proceed, we need the values of $k_q$ and $l_q$ on the horizon. According to Appendix~\ref{apd:metric}, they are
\begin{align}
    k_\mathrm{sGB}(2)=\frac{49}{80}\approx0.6,&\quad l_\mathrm{sGB}(2)=-\frac{63}{80}\approx-0.8,\\
    k_\mathrm{dCS}(2)=0,&\quad l_\mathrm{dCS}(2)=-\frac{709}{7168}\approx-0.1. \label{eq:gqg_cs}
\end{align}
Thus, the difference in the magnitude of the relative correction to the BZ power can be explained by the greater correction to the BH metric in the vicinity of the horizon in sGB than in dCS gravity. In fact, from Eqs.~\eqref{eq:prel_qg}--\eqref{eq:gqg_cs}, one recovers Eqs.~\eqref{eq:prel_gb}--\eqref{eq:prel_cs}.

With an expression of the BZ power in these quadratic theories, i.e., Eq.~\eqref{eq:power0_half}, one may wonder if measurements may be used to distinguish GR from these theories. As we will see, $\zeta_q$ and $\chi$ are degenerate to this order, so it is necessary to go to higher order, which we do next.

\subsection{The Blandford--Znajek Process in Quadratic Gravity to Second Relative Order in Spin}
\label{sec:qg_bz2}
We will now proceed to find the solution to the second relative order in spin. To this order, the stream Eq.~\eqref{eq:stream} now takes the form
\begin{align}
    L\psi_q^{(1,2)}=\psi_0\,s_q(x)\cos\theta\sin^2\theta,
\end{align}
where $s_q(x)$ is the radial source function, which is different for each theory. Considering the boundary conditions in Eqs.~\eqref{eq:flux_pole}--\eqref{eq:flux_horizon} and \eqref{eq:flux_infinity}, the solution then takes the form
\begin{align}
    \psi_q^{(1,2)}=\psi_0\,h_q(x)\cos\theta\sin^2\theta,
\end{align}
where $h_q(x)$ is the solution to the following inhomogeneous radial equation:
\begin{align}
    \frac{d}{dx}\left[\left(1-\frac{2}{x}\right)\frac{dh_q(x)}{dx}\right] - \frac{6h_q(x)}{x^2}=s_q(x), \label{eq:stream2_radial}
\end{align}
with the boundary conditions such that $h_q(x)$ is finite at $x=2$ and when $x\rightarrow\infty$. 
We have derived $s_q(x)$ and solved for $h_q(x)$ in closed-form. The expressions are rather long, and not illustrative, so we present them in Appendix~\ref{apd:test} (see Eqs.~\eqref{eq:radial_gb}--\eqref{eq:radial_sol_cs}). Here, we only summarize the behavior of the radial functions at the boundaries:
\begin{align}
    h_\mathrm{sGB}(2)=&-\frac{1865759261}{9408000}+\frac{11497\pi^2}{960}+\frac{49\pi^4}{60},\\
    h_\mathrm{dCS}(2)=&\frac{5562399}{40140800}-\frac{709\pi^2}{86016}
\end{align}
and
\begin{align}
    h_\mathrm{sGB}(x)\Big|_{x\rightarrow\infty}\sim&\frac{21}{80x}, \\
    h_\mathrm{dCS}(x)\Big|_{x\rightarrow\infty}\sim&-\frac{709}{28672x}.
\end{align}
Solving Eqs.~\eqref{eq:ipsi}--\eqref{eq:omegapsi} with the conditions of Eqs.~\eqref{eq:znajek_horizon}--\eqref{eq:znajek_infinity}, we find
\begin{align}
    I^{(1,3)}_q=&-2\pi\psi_0\bigg[\Omega^{(1,3)}_q\sin^2\theta +\Omega^{(1)}h_q(x)\sin^2\theta\cos^2\theta\notag\\
    &+\Omega_q^{(1,1)}f(x)\sin^2\theta\cos^2\theta\bigg],
\end{align}
and
\begin{align}
    \Omega^{(1,3)}_\mathrm{sGB}=&-\frac{21103}{403200M}
    +\bigg(\frac{626184387}{50176000}\notag\\&-\frac{11581\pi^2}{15360}-\frac{49 \pi ^4}{960}\bigg)\frac{\sin^2\theta}{M}, \label{eq:omega2_gb}\\
    \Omega^{(1,3)}_\mathrm{dCS}=&-\frac{169}{49152M}-\bigg(\frac{83313691}{5780275200}\notag\\&-\frac{709 \pi ^2}{688128}\bigg)\frac{\sin^2\theta}{M}.
    \label{eq:omega2_cs} 
\end{align}

The corrections to the BZ power of Eq.~\eqref{eq:power} at second relative order in spin are therefore
\begin{align}
    P^{(1,4)}_\mathrm{sGB}=&\pi\bigg(\frac{5652214483}{846720000}-\frac{2333\pi^2}{5760}-\frac{49\pi^4}{1800}\bigg)\frac{\psi_0^2}{M^2}, \label{eq:power2_gb}\\
    P^{(1,4)}_\mathrm{dCS}=&-\pi\left(\frac{163742291}{10838016000}-\frac{709\pi^2}{860160}\right)\frac{\psi_0^2}{M^2}. \label{eq:power2_cs}
\end{align}
Collecting all results so far, we have
\begin{widetext}
\begin{align}
    P_\mathrm{sGB}=&\left[\frac{\pi}{24}
    +\frac{7\pi}{80}\zeta_\mathrm{sGB}\right]\frac{\psi_0^2\chi^2}{M^2}
    +\left[\frac{\pi(56-3\pi^2)}{1080} +\pi\bigg(\frac{5652214483}{846720000}-\frac{2333\pi^2}{5760}-\frac{49\pi^4}{1800}\bigg)\zeta_\mathrm{sGB}\right]\frac{\psi_0^2\chi^4}{M^2}
    +\mathcal{O}(\zeta_\mathrm{sGB}^2,\chi^6),\\
    P_\mathrm{dCS}=&\left[\frac{\pi}{24}
    -\frac{709\pi}{86016}\zeta_\mathrm{dCS}\right]\frac{\psi_0^2\chi^2}{M^2}
    +\left[\frac{\pi(56-3\pi^2)}{1080}-\pi\left(\frac{163742291}{10838016000}-\frac{709\pi^2}{860160}\right)\zeta_\mathrm{dCS}\right]\frac{\psi_0^2\chi^4}{M^2}
    +\mathcal{O}(\zeta_\mathrm{dCS}^2,\chi^6).
\end{align}
For comparison, the horizon angular frequencies up to the same relative order are
\begin{align}
    \Omega_{\mathrm{H},\mathrm{sGB}}=&\left(\frac{1}{4}+\frac{21}{80}\zeta_\mathrm{sGB}\right)\frac{\chi}{M}
    +\left(\frac{1}{16} -\frac{21103}{201600}\zeta_\mathrm{sGB}\right)\frac{\chi^3}{M}
    +\mathcal{O}(\zeta_\mathrm{sGB}^2,\chi^5),\\
    \Omega_{\mathrm{H},\mathrm{dCS}}=&\left(\frac{1}{4}-\frac{709}{28672}\zeta_\mathrm{dCS}\right)\frac{\chi}{M}
    +\left(\frac{1}{16} -\frac{169}{24576}\zeta_\mathrm{dCS}\right)\frac{\chi^3}{M}
    +\mathcal{O}(\zeta_\mathrm{dCS}^2,\chi^5).
\end{align}
\end{widetext}
We see from these expressions that although $P_q \propto \Omega_{\mathrm{H},q}^2$ at leading order in $\chi$, this approximation breaks down at next-to-leading order. This is true in GR and in both sGB and dCS gravity.

\begin{figure}[htbp]
    \centering
    \includegraphics[width=0.48\textwidth]{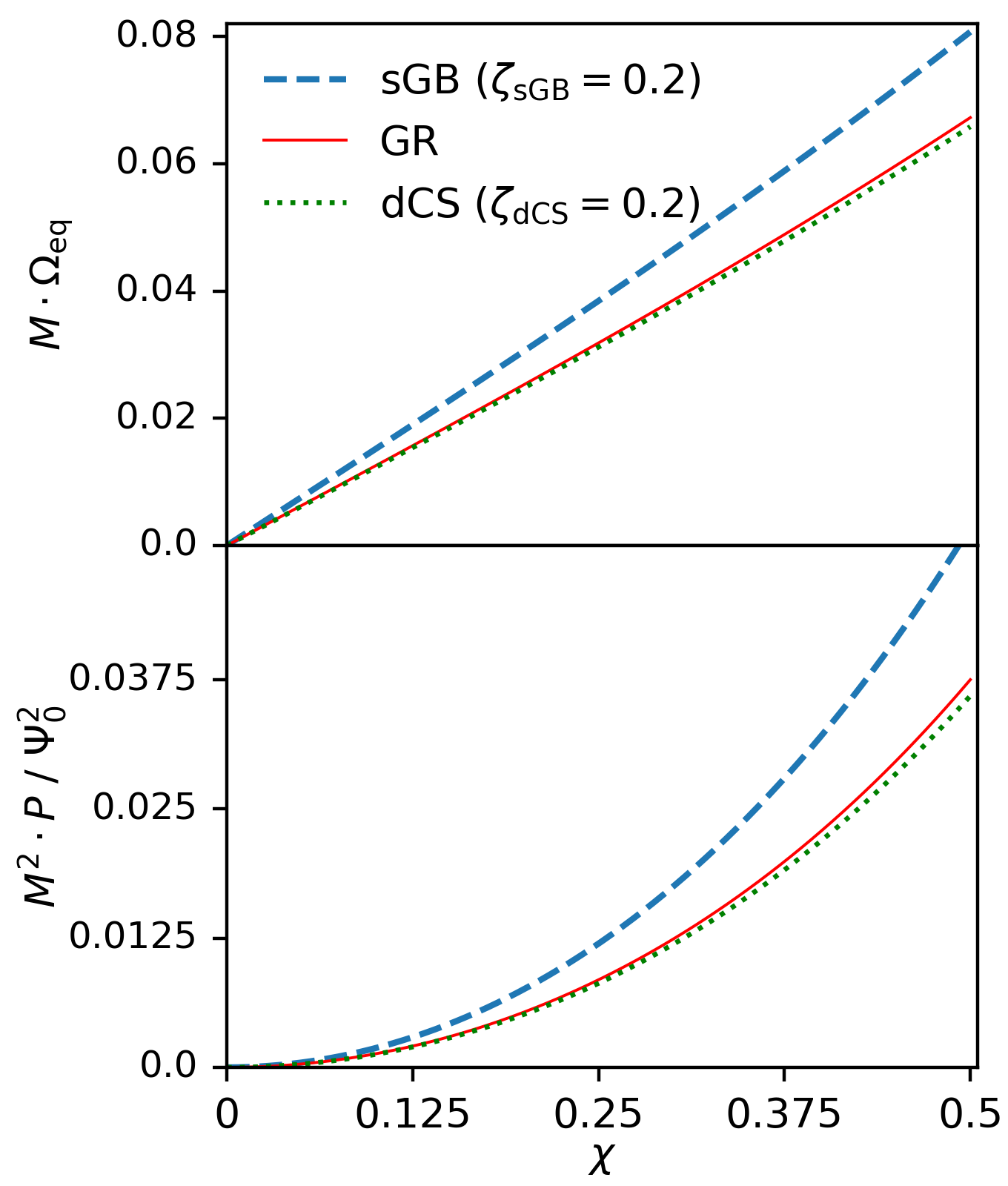}
    \caption{The rotation frequency of the EM field in the equatorial plane, $\Omega_\mathrm{eq}$, (top) and the BZ power, $P$, (bottom) as functions of the BH spin $\chi$ for GR (solid line), sGB (dashed lines), and dCS gravity (dotted lines), respectively. These quantities are computed up to second relative order in the small rotation approximation, i.e.,~$\mathcal{O}(\chi^3)$ for $\Omega_\mathrm{eq}$ and $\mathcal{O}(\chi^4)$ for $P$. In sGB and dCS gravity, the coupling constants $\zeta_\mathrm{sGB}$ and $\zeta_\mathrm{dCS}$ are both set to $0.2$ for illustrative purposes, and their modifications to GR are considered to first order in their coupling constants. Deviations from the GR result are larger in sGB than in dCS gravity, as expected.}
    \label{fig:freq_param}
\end{figure}

Figure~\ref{fig:freq_param} shows the equatorial rotation frequency $\Omega_\mathrm{eq}\equiv\Omega(\theta=\pi/2)$ and the BZ power $P$ as functions of the BH spin $\chi$, up to second relative order. As found to leading order in the previous section, the BZ power is enhanced in sGB and quenched in dCS, with respect to the prediction of GR. As these solutions are only valid in the small-coupling approximation, we have fixed the dimensionless coupling constants $\zeta_q=0.2$ to qualitatively show the different behaviors of the BZ power.

As we have only considered solutions up to second relative order in spin, it was unnecessary to follow the procedure presented by Armas et al.~\cite{Armas:2020mio}, i.e., matched asymptotics plus smoothness checks. Even though the results presented in~\cite{Armas:2020mio} were derived within GR, we expected a similar behaviour of the BZ solution in these modified theories. However, as the BH metrics in sGB and dCS gravity are only known in the mid-region, a rigorous proof of this behaviour cannot be provided, as we explain in detail in Appendix~\ref{apd:test}. Despite that, we have applied the method proposed by Armas et al. using resummed metrics for sGB and dCS and found the field solutions in the near and far expansions are trivial, and that the smoothness assumption holds up to second relative order in the spin. Since our resummation recovers the exact Kerr metric and shifts the coordinate singularity to the modified horizon, we argue that this resummation is likely to work in the entire domain. A detailed presentation of these calculations is presented in Appendix~\ref{apd:test}.

\section{Astrophysical Implications}
\label{sec:implications}

The BZ process has three free parameters\footnote{There will naturally be more degrees of freedom when considering other configurations or symmetries of the disk and jet than those considered in this work (for instance, see~\cite{1977MNRAS.179..433B,1985MNRAS.212..899B,Gralla:2015vta}). For example, state-of-the-art GRMHD models can display a  jet--disk boundary layer that fluctuates strongly, and therefore more parameters may be needed to describe the jet power~\cite{Wong:2021qbe}.}: the angular velocity of the event horizon ($\Omega_\mathrm{H}$, which only depends on the BH's parameters), the rotation frequency of magnetic field lines ($\Omega$, which is dictated by the dynamics of the system), and the magnetic flux through the horizon ($\psi_0$). Therefore, measurements of \emph{only} the jet power cannot be used to learn about the underlying physics of the process. Within GR, it is customary to assume $\Omega=\Omega_\mathrm{H}/2$ or to check for a square proportionality of the jet power with $\Omega_\mathrm{H}$ when fitting data~\cite{Steiner:2012ap,Blandford:2018iot,Chen:2021ffu}. Even within GR, a clear observational signature of the BZ mechanism is still missing, although it may be possible that future observations may provide the quality and type of data necessary. 

Pei et al.,~\cite{Pei:2016kka}, assuming $\Omega=\Omega_\mathrm{H}/2$, combined estimates of the jet power with \emph{independent} measurements of the black hole spin and found that current data cannot place informative constraints on the metric deformation parameters. However, in the presence of better measurements, they conjectured that such types of tests may be possible. Given this, let us now hypothesize about tests of gravity in the future, i.e., if, for example, $\Omega$ can be measured and \emph{independent} high quality measurements of the BH's spin become possible. Would high quality data be able to distinguish GR from other theories of gravity using the BZ power? As we will show below, in addition to precise future measurements, a magnetospheric solution that goes beyond second order will also be required.

Let us assume $\Omega\approx\Omega_\mathrm{H}/2$ to write Eq.~\eqref{eq:power0_half} as
\begin{equation}
    P_q(\zeta_q,\chi)=\frac{2\pi}{3}\psi_0^2\left[\Omega_q(\zeta_q,\chi)\right]^2+\mathcal{O}(\zeta_q^2,\chi^4). \label{eq:power0_omega0}
\end{equation}
From this expression, one can see that $P_q$ is a function that only depends on $\Omega_q$ at leading order in spin. This implies that, to this order, $\zeta_q$ and $\chi$ are degenerate. In other words, we will not be able to determine \emph{both} the coupling constant $\zeta_q$ and the spin $\chi$ even if \emph{both} the BZ power $P_q$ and the field rotation frequency $\Omega_q$ are measured. Note that Eq.~\eqref{eq:power0_omega0} holds as long as the magnetosphere dynamics maximizes the BZ power, and therefore, this degeneracy is a general issue under such a condition.

To higher order in spin, however, this is not the case. To see whether the degeneracy breaks between $\zeta_q$ and $\chi$, we vary $\zeta_q\rightarrow\zeta_q+\delta\zeta_q$ and $\chi\rightarrow\chi+\delta\chi$ and study the following Jacobian determinant:
\begin{equation}
    \left|\frac{\delta\ln(P,\Omega)}{\delta\ln(\zeta,\chi)}\right|_q
    \equiv
    \begin{vmatrix}
    \delta\ln P_q/\delta\ln\zeta_q & \delta\ln P_q/\delta\ln\chi \\
    \delta\ln\Omega_q/\delta\ln\zeta_q & \delta\ln\Omega_q/\delta\ln\chi
    \end{vmatrix}. \label{eq:def_jacobian}
\end{equation}
Evaluating Eq.~\eqref{eq:def_jacobian} with $P_q$ and $\Omega_q$ to leading order in spin, this Jacobian vanishes, and thus $\zeta_q$ and $\chi$ are degenerate at leading order in spin as mentioned above. Now if we add the corrections at second relative order in spin, as given in Eqs.~\eqref{eq:iomega2_gr}--\eqref{eq:power_gr} and Eqs.~\eqref{eq:omega2_gb}--\eqref{eq:power2_cs}, one finds 
\begin{align}
    \left|\frac{\delta\ln(P,\Omega)}{\delta\ln(\zeta,\chi)}\right|_\mathrm{sGB}=&\left(-\frac{616991987}{31360000}+\frac{11329 \pi^2}{9600}+\frac{49\pi^4}{600}\right)\notag\\
    &\times(3+5\cos2\theta)\,\zeta_\mathrm{sGB}\chi^2\notag\\&+\mathcal{O}(\zeta_\mathrm{sGB}^2,\chi^4), \\
    \left|\frac{\delta\ln(P,\Omega)}{\delta\ln(\zeta,\chi)}\right|_\mathrm{dCS}=&\left(-\frac{16442609}{3612672000}+\frac{709\pi^2}{860160}\right)\notag\\
    &\times(3+5\cos2\theta)\,\zeta_\mathrm{dCS}\chi^2\notag\\&+\mathcal{O}(\zeta_\mathrm{dCS}^2,\chi^4).
\end{align}
Therefore the degeneracy between $\zeta_q$ and $\chi$ breaks when the BZ power to second relative order in spin is considered. Given that the degeneracy only breaks at higher orders in the slow-rotation approximation, we expect that a determination of or constraint on $\zeta_q$ and $\chi$ by measuring $P_q$ and $\Omega_q$ will only be possible for rapidly-rotating BHs, provided that \emph{both} quantities are computed accurately. 

\section{Discussion}
\label{sec:discussion}

We have studied the BZ process in two well-motivated quadratic gravity theories: sGB and dCS gravity. We solved the BH magnetosphere analytically to first order in the small-coupling approximation and to second relative order in the slow-rotation approximation, assuming a split-monopole configuration. We found that the power of energy extraction from the BH, compared to the predictions of GR, is enhanced in sGB gravity and quenched in dCS gravity. 

We have further shown that, for these quadratic BH solutions, the strategy to solve for the fields proposed by Armas et al.~\cite{Armas:2020mio} cannot be applied, as the approximated BH solutions do not fit into a matched asymptotics framework. However, as shown by Armas et al.~\cite{Armas:2020mio}, in GR, the inclusion of the condition in Eq.~\eqref{eq:znajek_infinity} is sufficient for solving the BZ process up to second relative order in the slow-rotation approximation, and the matched asymptotics and the smoothness issue can be neglected. By studying a resummed version of the quadratic gravity BH solutions, we have argued that the same holds true in quadratic gravity.

Previous studies of the BZ mechanism outside GR~\cite{Pei:2016kka,Konoplya:2021qll,Banerjee:2020ubc} have only been considered to first relative order in the small-spin expansion, where a degeneracy occurs that hinders our ability to use this mechanism to distinguish GR from other theories of gravity. Furthermore,~\cite{Pei:2016kka,Konoplya:2021qll} have used parametrically deformed metrics with only one deformation parameter. However, most of the known modified solutions cannot be mapped to such metrics (with only one deformation parameter), and when multiple parameters are included in the analyses of observables, the degeneracies between the astrophysical and BH parameters are enhanced, making theory-agnostic studies very challenging~\cite{Cardenas-Avendano:2019pec,Volkel:2020xlc}. Therefore studies of specific theories, as the one presented here or in~\cite{Banerjee:2020ubc}, should be seen as complementary.
   
Our results motivate further analytical and numerical studies of the BZ process in modified theories of gravity and continue to pave the road towards addressing whether the phenomena related to the BZ mechanism can be used to learn about fundamental physics from BH observations.    

\acknowledgments
We thank Dimitry Ayzenberg, Samuel Gralla, and George Wong for useful discussions and comments. Y.X., N.Y., and C.F.G. were supported by NSF grant 20-07936. A.C.-A. acknowledges funding from Will and Kacie Snellings, and from Fundaci\'on Universitaria Konrad Lorenz (Project 5INV1). 
\appendix
\section{Slow-Rotation, Small-Coupling Black Hole Solutions in Quadratic Gravity}
\label{apd:resum}

This Appendix explicitly shows the transformation of coordinates from Hartle--Thorne to Boyer--Lindquist coordinates and the resummed metrics used in the main text. 

\subsection{Coordinate transformation from Hartle--Thorne to Boyer--Lindquist coordinates}
\label{apd:metric}

The BH solutions used in this work were derived in Hartle--Thorne (HT) coordinates in ~\cite{Maselli:2015tta,Maselli:2017kic} for sGB and dCS gravity, respectively. Below we show explicitly, up to $\mathcal{O}(\zeta_q,\chi^5)$, the transformation from HT coordinates, i.e., $\left(t,r_\mathrm{HT},\theta_{\mathrm{HT}},\phi\right)$, to BL coordinates, i.e., $\left(t,r,\theta,\phi\right)$. The transformation is assumed to be of the form

\begin{align}
r_\mathrm{HT,q}=\sum^n\left(r_{\mathrm{HT}}^{\left(n\right)}+\zeta_{q}r_{\mathrm{HT,q}}^{\left(n\right)}\left[r_{\mathrm{BL}},\theta_{\mathrm{BL}}\right]\right)\chi^{n},\\
\theta_\mathrm{HT,q}=\sum^n\left(\theta_{\mathrm{HT}}^{\left(n\right)}+\zeta_{q}\theta_{\mathrm{HT,q}}^{\left(n\right)}\left[r_{\mathrm{BL}},\theta_{\mathrm{BL}}\right]\right)\chi^{n},
\end{align}
where the integer $(n)$ stands for the $n$th order in the spin $\chi$. Using this ansatz, the transformation $g_{\mu\nu}^{\mathrm{BL}}=\Lambda_{\mu}^{\alpha}\Lambda_{\nu}^{\beta}g_{\alpha\beta}^{HT}$, with $\Lambda_{\mu}^{\alpha}=\partial x_{\mathrm{HT}}^{\alpha}/\partial x_{\mathrm{BL}}^{\mu}$, is solved order by order. Starting with the GR terms, the transformation requires only to solve algebraic equations because the Kerr solution is known in both coordinate systems. In particular, it is enough to apply the transformation and simultaneously solve for $r_{\mathrm{HT}}^{\left(n\right)}\left[r_{\mathrm{BL}},\theta_{\mathrm{BL}}\right]$ and $\theta_{\mathrm{HT}}^{\left(n\right)}\left[r_{\mathrm{BL}},\theta_{\mathrm{BL}}\right]$ in $g_{tt}^{\mathrm{BL}}-g_{tt}^{\mathrm{HT}}=0$ and $g_{\phi\phi}^{\mathrm{BL}}-g_{\phi \phi}^{\mathrm{HT}}=0$, order by order. 

This exact procedure also applies to both sGB and dCS, but the equations start to be coupled partial differential equations,instead of algebraic, for $n\geq3$, as the solutions were only previously known in BL up to second order in the spin~\cite{,Yagi:2012ya,Ayzenberg:2016ynm}. Thus, one solves, order by order, for $r_{\mathrm{q,HT}}^{\left(n\right)}\left[r_{\mathrm{BL}},\theta_{\mathrm{BL}}\right]$ and $\theta_{\mathrm{q,HT}}^{\left(n\right)}\left[r_{\mathrm{BL}},\theta_{\mathrm{BL}}\right]$ in the resulting coupled partial differential equations. For simplicity, we require that our transformation satisfies $g_{r\theta}=0$. The explicit resulting coordinate transformation we used in this work is:
\begin{widetext}
\begin{align}
    r_\mathrm{HT,sGB}=&r_\mathrm{HT,GR}
    -\zeta_\mathrm{sGB}\chi^2\frac{M^4}{12r^3}\left(1+\frac{4 M}{r}+\frac{61 M^2}{3 r^2}+\frac{54 M^3}{r^3}+\frac{46 M^4}{5 r^4}-\frac{1696 M^5}{15 r^5}-\frac{368 M^6}{r^6}\right)(1+3\cos2\theta) \notag\\
    &+\zeta_\mathrm{sGB}\chi^4\frac{M^4}{8r^3}\Bigg[\bigg(1+\frac{4 M}{r}+\frac{34606 M^2}{2625 r^2}+\frac{19556 M^3}{525 r^3}+\frac{8017663 M^4}{55125 r^4}+\frac{322582 M^5}{875 r^5}+\frac{194692 M^6}{525 r^6}\notag\\
    &-\frac{290140 M^7}{441 r^7}-\frac{515756 M^8}{105 r^8}+\frac{4608 M^9}{5 r^9}-\frac{11552 M^{10}}{r^{10}}\bigg)\cos2\theta
    -\frac{3019M^2}{1750r^2}\bigg(1+\frac{14220 M}{3019 r}-\frac{2811413 M^2}{63399 r^2}\notag\\
    &-\frac{101488 M^3}{9057 r^3}+\frac{372990 M^4}{3019 r^4}-\frac{18494900 M^5}{63399 r^5}-\frac{639400 M^6}{9057 r^6}-\frac{10197600 M^7}{3019 r^7}+\frac{25816000 M^8}{3019 r^8}\bigg)\cos^22\theta\Bigg],\\
    \theta_\mathrm{HT,sGB}=&\theta_\mathrm{HT,GR},
\end{align}
and 
\begin{align}
    r_\mathrm{HT,dCS}=&r_\mathrm{HT,GR}
    -\zeta_\mathrm{dCS}\chi^4\frac{661M^6}{43000r^5} \Bigg[\bigg(1+\frac{4005 M}{661 r}+\frac{215826 M^2}{4627 r^2}+\frac{175636 M^3}{661 r^3}+\frac{343404 M^4}{661 r^4}-\frac{829404 M^5}{4627 r^5} \notag\\
    &-\frac{1532520 M^6}{661 r^6}-\frac{2467584 M^7}{661 r^7}\bigg)\cos2\theta
    -\frac{117}{1322}\bigg(1+\frac{5 M}{r}+\frac{143834 M^2}{273 r^2}-\frac{12676 M^3}{39 r^3}-\frac{78380 M^4}{13 r^4}\notag\\
    &-\frac{690876 M^5}{91 r^5}+\frac{20952 M^6}{13 r^6}+\frac{822528 M^7}{13 r^7}\bigg)\cos^22\theta
    \Bigg],\\
    \theta_\mathrm{HT,dCS}=&\theta_\mathrm{HT,GR},
\end{align}
where the transformations in GR are given by
\begin{align}
    r_\mathrm{HT,GR}=&r
    -\chi^2\frac{M^2}{4r} \left(1+\frac{M}{r}-\frac{6 M^2}{r^2}\right)\cos2\theta -\chi^4\frac{M^4}{8r^3} \Bigg[1+\frac{3M}{r}-\frac{36M^2}{5r^2}-\frac{72M^3}{5r^3}+\frac{8M^4}{5r^4}\notag\\
    &-2\left(1+\frac{3 M}{r}-\frac{18 M^2}{r^2}-\frac{42 M^3}{r^3}+\frac{36 M^4}{r^4}\right)\cos^2\theta
    +\left(1+\frac{3 M}{r}-\frac{28 M^2}{r^2}-\frac{60 M^3}{r^3}+\frac{192 M^4}{r^4}\right)\cos^4\theta\Bigg],\\
    \theta_\mathrm{HT,GR}=&\theta +\chi^2\frac{M^2}{4r^2}\left(1+\frac{2 M}{r}\right)\sin2\theta \notag\\
    &-\chi^4\frac{M^4}{8r^4}\Bigg[\left(1+\frac{4 M}{r}+\frac{5 M^2}{r^2}+\frac{6 M^3}{r^3}\right)\sin2\theta
    -\frac{1}{4}\left(1+\frac{4 M}{r}+\frac{2 M^2}{r^2}-\frac{12 M^3}{r^3}\right)\sin4\theta\Bigg].
\end{align}
\end{widetext}

The resulting metric expressions in BL coordinates are available in a Mathematica notebook provided in the Supplemental Material.

\subsection{Resummation of Slow-Rotation, Small-Coupling Black Hole Solutions}

As discussed in the main text, it is suitable to re-express the metric solutions as a resummation such that analytic calculations, like the one presented in Appendix~\ref{apd:test}, can be performed. In particular, our resummation will provide a metric with the following properties:
\begin{enumerate}[(i)]
    \item differs from the series-expanded metric only by terms of $\mathcal{O}(\zeta_q^2,\chi^6)$,
    \item recovers the exact Kerr metric when taking $\zeta_q\rightarrow0$, 
    \item encodes the location of the corrected horizon $r=r_{\mathrm{H},q}$ (not at $r=2M$) through a redefinition of the $\Delta$ function of the Kerr metric,
    \item encodes the location of the corrected ergosphere $r=r_{\mathrm{ergo},q}$ through a redefinition of the $\Sigma$ function of the Kerr metric,
    \item avoids introducing naked singularities or closed time-like curves.
\end{enumerate}
Indeed, item (i) must hold for any resummation procedure (almost by definition of what we mean by resummation). Items (ii)--(v), however, are additional requirements we impose to refine our resummation procedure, but even then, this scheme is still not unique. 

Given a series-expanded solution to higher order than  $\mathcal{O}(\zeta_q,\chi^5)$, one can repeat this procedure to get more accurate representations of the solution. 

Let us first consider the coordinate singularity. Yagi et al.~\cite{Yagi:2012ya} have proposed a resummation strategy that shifts the coordinate singularity in the approximate dCS BH solution from $r=2M$ to $r=r_\mathrm{H,dCS}$. This resummation strategy works by taking $\Delta\rightarrow\Delta_\mathrm{dCS}$ in the Kerr piece of $g^\mathrm{dCS}_{rr}$ and taking $(r-2M)\rightarrow (r-r_\mathrm{H,dCS})$ in the dCS modification piece of $g_{rr}$. 
Here, $\Delta_\mathrm{dCS}$ deviates from $\Delta$ in a way such that $\Delta_\mathrm{dCS}=0$ occurs for $r=r_\mathrm{H,dCS}$. Ayzenberg and Yunes~\cite{Ayzenberg:2018jip} (there is a typo in their expressions that we correct here) have computed $\Delta_\mathrm{dCS}$ to $\mathcal{O}(\zeta_q,\chi^5)$ :

\begin{align}
    \Delta_\mathrm{dCS}=\Delta+M^2\zeta_\mathrm{dCS}\left(\frac{915}{14336}\chi^2+\frac{131879}{6881280}\chi^4\right). \label{eq:resum_delta_cs}
\end{align}
Using this transformation, $\tilde{g}^\mathrm{dCS}_{rr}\equiv g^\mathrm{dCS}_{rr}\Delta_\mathrm{dCS}$ does not become singular at $r=2M$ when evaluated up to $\mathcal{O}(\zeta_\mathrm{dCS},\chi^5)$. Therefore, we can apply a simpler resummation strategy by just computing $\tilde{g}^\mathrm{dCS}_{rr}$ up to $\mathcal{O}(\zeta_\mathrm{dCS},\chi^5)$ and replacing
\begin{align}
    g^\mathrm{dCS}_{rr}\rightarrow\tilde{g}^\mathrm{dCS}_{rr}/\Delta_\mathrm{dCS}.
\end{align}
The same procedure also applies in sGB, and therefore
\begin{align}
    \Delta_\mathrm{sGB}=\Delta+M^2\zeta_\mathrm{sGB}\left(\frac{49}{20}-\frac{311}{480}\chi^2-\frac{813569}{1612800}\chi^4\right). \label{eq:resum_delta_gb}
\end{align}

The next step is to make sure that we recover the exact Kerr metric when taking $\zeta_q\rightarrow0$. Here, we consider replacing terms that appear as $1/r^{n}~(n>0)$ with $(r/\Sigma_q)^{n}$, where $\Sigma_q$ deviates from $\Sigma$ in a way such that $\Sigma_q-2Mr=0$ gives the correct value of the ergosphere $r_{\mathrm{ergo},q}(\theta)$. The results are
\begin{align}
    \Sigma_\mathrm{sGB}=&\Sigma + M^2\zeta_\mathrm{sGB}\bigg[\frac{49}{20} - \left(\frac{191}{160}+\frac{131}{240}\cos^2\theta\right)\chi^2 \notag\\
    &+ \bigg(\frac{14370073}{56448000}+\frac{4829219}{1764000}\cos^2\theta\notag\\
    &-\frac{16448333}{4704000}\cos^4\theta\bigg)\chi^4\bigg], \label{eq:resum_sigma_gb}\\
    \Sigma_\mathrm{dCS}=&\Sigma + M^2\zeta_\mathrm{dCS}\bigg[\left(\frac{3751}{14336}-\frac{709}{3584}\cos^2\theta\right)\chi^2 \notag\\
    &-\bigg(\frac{1922747}{48168960}+\frac{34351}{150528}\cos^2\theta\notag\\
    &-\frac{230637}{802816}\cos^4\theta\bigg)\chi^4\bigg]. \label{eq:resum_sigma_cs}
\end{align}
We note that we do not replace all $1/r^{n}$ terms at the same time; otherwise, the exact Kerr metric cannot be recovered in the GR sector. Instead, we order the replacement as follows. Given a metric component $g^q_{\mu\nu}$ in the original BH solution, we calculate its Laurent expansion about $r=0$. The result should take the following form:
\begin{align}
    g^q_{\mu\nu}=\sum_{n=0}^{N_+}C_nr^n + \sum_{n=1}^{N_-}D^{(0)}_n/r^{n},
\end{align}
where $N_+$ and $N_-$ are finite non-negative integers, and $C_n$ and $D_n$ are precise up to $\mathcal{O}(\zeta_q,\chi^5)$.
The first sum is non-diverging, while the second sum contains all diverging terms that has to be replaced. We first take
\begin{align}
    D^{(0)}_1/r\rightarrow D^{(0)}_1r/\Sigma_q.
\end{align}
Now $D^{(0)}_1r/\Sigma_q$ is non-diverging. We can then rewrite $g^q_{\mu\nu}$ as follows:
\begin{align}
    g^q_{\mu\nu}=\left[\sum_{n=0}^{N_+}C_nr^n + D^{(0)}_1r/\Sigma_q\right] + \sum_{n=2}^{N_-}D^{(1)}_n/r^{n},
\end{align}
where we have put all non-diverging terms in the bracket and adjusted the diverging terms to keep $g^q_{\mu\nu}$ precise up to $\mathcal{O}(\zeta_q,\chi^5)$. At the $i$th step, we replace
\begin{align}
    D^{(i-1)}_i/r^i\rightarrow D^{(i-1)}_i(r/\Sigma_q)^i,
\end{align}
and rewrite
\begin{align}
    g^q_{\mu\nu}=&\left[\sum_{n=0}^{N_+}C_nr^n + \sum_{n=1}^iD^{(n-1)}_n(r/\Sigma_q)^n\right] \notag\\&+ \sum_{n=i+1}^{N_-}D^{(i)}_n/r^{n},
\end{align}
where each $D^{(i)}_n$ is adjusted from $D^{(i-1)}_n$ so that the above expression holds up to $\mathcal{O}(\zeta_q,\chi^5)$.
By the $N_-$th step, there should be nothing left for the diverging part, and the whole replacement is completed. We have checked that the obtained resummed metrics recover the exact Kerr metric when taking $\zeta_q\rightarrow0$, and they recover the series-expanded metrics when replacing $\Delta_q$ and $\Sigma_q$ using Eqs.~\eqref{eq:resum_delta_cs}, \eqref{eq:resum_delta_gb}, and \eqref{eq:resum_sigma_gb}--\eqref{eq:resum_sigma_gb} and re-expanding to $\mathcal{O}(\zeta_q,\chi^5)$.

The result of this procedure gives the following resummed BH solutions, which we only show here up to $\mathrm{O}(\zeta_q,\chi^2)$:
\begin{widetext}
\begin{align}
    g^\mathrm{sGB}_{tt}=&\left(-1+\frac{2Mr}{\Sigma_\mathrm{sGB}}\right) \left[1- \zeta_\mathrm{sGB}\frac{137M^3r^3}{30\Sigma_\mathrm{sGB}^3}\left(1+\frac{14 M r}{137\Sigma_\mathrm{sGB}}-\frac{104 M^2 r^2}{137\Sigma_\mathrm{sGB}^2}-\frac{400 M^3 r^3}{137\Sigma_\mathrm{sGB}^3}\right)\right], \\
    g^\mathrm{sGB}_{rr}=&\frac{1}{\Delta_\mathrm{sGB}}\left[r^2+\chi^2M^2\cos^2\theta +\zeta_\mathrm{sGB}\frac{29M^2}{20}\left(1+\frac{38 M r}{29\Sigma_\mathrm{sGB}}-\frac{28 M^2 r^2}{3\Sigma_\mathrm{sGB}^2}-\frac{1744 M^3 r^3}{87\Sigma_\mathrm{sGB}^3}-\frac{3680 M^4 r^4}{87\Sigma_\mathrm{sGB}^4}\right)\right],\\
    g^\mathrm{sGB}_{\theta\theta}=&r^2+\chi^2M^2\cos^2\theta,\\
    g^\mathrm{sGB}_{\phi\phi}=&r^2\sin^2\theta+\chi^2M^2\left(1+\frac{2Mr}{\Sigma_\mathrm{sGB}}\sin^2\theta\right)\sin^2\theta,\\
    g^\mathrm{sGB}_{t\phi}=&-\chi\frac{2Mr}{\Sigma_\mathrm{sGB}}\sin^2\theta
    -\zeta_\mathrm{sGB}\frac{43M^4r^3}{10\Sigma_\mathrm{sGB}^3}\left(1 -\frac{280 M r}{129\Sigma_\mathrm{sGB}}-\frac{60 M^2 r^2}{43\Sigma_\mathrm{sGB}^2}-\frac{96 M^3 r^3}{43\Sigma_\mathrm{sGB}^3}+\frac{800 M^4 r^4}{129\Sigma_\mathrm{sGB}^4}\right)\sin^2\theta.
\end{align}
\begin{align}
    g^\mathrm{dCS}_{tt}=&-1+\frac{2Mr}{\Sigma_\mathrm{dCS}},\\
    g^\mathrm{dCS}_{rr}=&\frac{1}{\Delta_\mathrm{dCS}}\left(r^2+\chi^2M^2\cos^2\theta\right),\\
    g^\mathrm{dCS}_{\theta\theta}=&r^2+\chi^2M^2\cos^2\theta,\\
    g^\mathrm{dCS}_{\phi\phi}=&r^2\sin^2\theta+\chi^2M^2\left(1+\frac{2Mr}{\Sigma_\mathrm{dCS}}\sin^2\theta\right)\sin^2\theta,\\
    g^\mathrm{dCS}_{t\phi}=&-\chi\frac{2Mr}{\Sigma_\mathrm{dCS}}\sin^2\theta
    +\zeta_\mathrm{dCS}\chi\frac{5M^5r^4}{\Sigma_\mathrm{dCS}^4}\left(1+\frac{12 M r}{7 \Sigma_\mathrm{dCS}}+\frac{27 M^2 r^2}{10 \Sigma_\mathrm{dCS}^2}\right)\sin^2\theta,
\end{align}
\end{widetext}
The complete expressions of the resummed metric up to $\mathcal{O}(\zeta_q,\chi^5)$ are available in a Mathematica notebook provided in the Supplemental Material.

\section{Blandford--Znajek Solution in Quadratic Gravity Using Matched Asymptotics}
\label{apd:test}

In Sec.~\ref{sec:qg_bz0}--\ref{sec:qg_bz2}, we derived the BZ process following a similar procedure as shown in e.g.,~\cite{1977MNRAS.179..433B,McKinney_2004}, but we adopted the boundary conditions presented by Armas et al.~\cite{Armas:2020mio}. In this appendix, we present the solutions to the BZ mechanism in quadratic gravity following the procedure presented by Armas et al.~\cite{Armas:2020mio} and show that the results coincide. 

We start by defining three distinctive slow-rotation expansions, namely ``near,'' ``mid'' and ``far,'' by their length scales, $R_\mathrm{near}\ll R_\mathrm{mid}\ll R_\mathrm{far}$, where:
\begin{align}
    R_\mathrm{near}=&a^2/M, \label{eq:scale_near}\\
    R_\mathrm{mid}=&M, \label{eq:scale_mid}\\
    R_\mathrm{far}=&M^2/a. \label{eq:scale_far}
\end{align}
The mass and the spin are, accordingly, now expressed as:
\begin{align}
    M=&R_\mathrm{near}\chi^{-2}=R_\mathrm{mid}=R_\mathrm{far}\chi, \label{eq:mass_scale}\\
    a=&R_\mathrm{near}\chi^{-1}=R_\mathrm{mid}\chi=R_\mathrm{far}\chi^2. \label{eq:spin_scale}
\end{align}
Analogously, the $r$ coordinate should also be replaced by the following dimensionless radii:
\begin{align}
    y=&(r-r_\mathrm{H})/R_\mathrm{near},\\
    x=&r/R_\mathrm{mid},\\
    \bar{x}=&r/R_\mathrm{far}. \label{eq:radius_far}
\end{align}
Let $Q_\mathrm{near}(y)$, $Q_\mathrm{mid}(x)$, and $Q_\mathrm{far}(\bar{x})$ be some field variables in the three different expansions. The boundary conditions on the horizon and at infinity should apply to $Q_\mathrm{near}|_{y=0}$ and $Q_\mathrm{far}|_{\bar{x}\rightarrow\infty}$, respectively.
In addition, matched asymptotics requires that
\begin{align}
    Q_\mathrm{near}\big|_{y\rightarrow\infty}\sim &Q_\mathrm{mid}\big|_{\bar{x}\rightarrow 2},\\
    Q_\mathrm{mid}\big|_{x\rightarrow\infty}\sim &Q_\mathrm{far}\big|_{\bar{x}\rightarrow 0}.
\end{align}
For example, consider a term in the mid expansion that has the following dependence on $x$ in the vicinity of $x\rightarrow\infty$:
\begin{align}
    Q_\mathrm{mid}^{(4)}\big|_{x\rightarrow\infty}=R_\mathrm{mid}\left(x+\frac{1}{x}\right)+\cdots,
\end{align}
where ``$\cdots$'' means there could be other dependencies on $x$. 
In the vicinity of $\bar{x}\rightarrow0$, using $R_\mathrm{mid}=\chi R_\mathrm{far}$ and $x=\bar{x}/\chi$, one finds that
\begin{align}
    Q_\mathrm{far}^{(4)}\big|_{\bar{x}\rightarrow0}=&R_\mathrm{far}\bar{x}+\cdots,\\
    Q_\mathrm{far}^{(6)}\big|_{\bar{x}\rightarrow0}=&\frac{R_\mathrm{far}}{\bar{x}}+\cdots.
\end{align}

Given the characteristics of the three expansions in Eqs.~\eqref{eq:scale_near}--\eqref{eq:radius_far}, we recognize that the mid expansion coincides with the slow-rotation approximation presented above. As expected, the quadratic gravity metric solutions presented in Appendix~\ref{apd:metric} are given as mid expansions. In order to conduct the full procedure by Armas et al., we also need the metric solutions in the near and far expansions. 

We note that the far-expansion metric can be converted from the mid-expansion metric by replacing $M\rightarrow R_\mathrm{far}\chi$, $a\rightarrow R_\mathrm{far}\chi^2$, and $r\rightarrow R_\mathrm{far}\bar{x}$. On the other hand, for the near expansion, the same strategy is not guaranteed to work because negative powers will be involved when taking $M\rightarrow R_\mathrm{near}\chi^{-2}$ and $a\rightarrow R_\mathrm{near}\chi^{-1}$. In addition, the $r\rightarrow r_\mathrm{H}+R_\mathrm{near}y$ replacement also requires the metric to be well-defined near the horizon. This is why we have resummed the metric solutions in Appendix~\ref{apd:resum} such that the exact Kerr solution is recovered when $\zeta\rightarrow0$, and the coordinate singularity at $r=2M$ is shifted to the horizon radius $r_{\mathrm{H},q}$. 

Like in the main text, we consider up to second relative order in spin. We start by writing the GR solution found in~\cite{Armas:2020mio}. To leading order, it is
\begin{align}
    \psi_\mathrm{near}^{(0)}=&\psi_\mathrm{mid}^{(0)}=\psi_\mathrm{far}^{(0)}=\psi_0(1-\cos\theta), \label{eq:psi0_mid_gr} \\
    \chi^3 I_\mathrm{near}^{(3)}=&\chi I_\mathrm{mid}^{(1)}=I_\mathrm{far}^{(0)}=-\frac{2\pi\psi_0 a}{M^2}\omega_0\sin^2\theta,\\
    \chi^3 \Omega_\mathrm{near}^{(3)}=&\chi \Omega_\mathrm{mid}^{(1)}=\Omega_\mathrm{far}^{(0)}=\frac{a}{M^2}\omega_0,
\end{align}
where
\begin{align}
    \omega_0=\frac{1}{8}.\label{eq:omega0_sub_gr}
\end{align}
Note that because $I$ and $\Omega$ are proportional to $a/M^2$, their scaling behavior with respect to $\chi$ varies in different expansions according to Eqs.~\eqref{eq:mass_scale}--\eqref{eq:spin_scale}. 

At first relative order,
\begin{align}
    \psi_\mathrm{near}^{(1)}=&\psi_\mathrm{mid}^{(1)}=\psi_\mathrm{far}^{(1)}=0,\\
    \chi^3 I_\mathrm{near}^{(4)}=&\chi I_\mathrm{mid}^{(2)}=I_\mathrm{far}^{(1)}=0,\\
    \chi^3 \Omega_\mathrm{near}^{(4)}=&\chi \Omega_\mathrm{mid}^{(2)}=\Omega_\mathrm{far}^{(1)}=0,
\end{align}
while to second relative order, the mid expansion is
\begin{align}
    \psi_\mathrm{mid}^{(2)}=&\psi_0 f(x)\sin^2\theta\cos\theta,\\
    I_\mathrm{mid}^{(3)}=&-\frac{2\pi\psi_0}{M}\sin^2\theta\left[\omega_2(\theta)+\frac{1}{4}f(x)\cos^2\theta\right],\\
    \Omega_\mathrm{mid}^{(3)}=&\frac{1}{M}\omega_2(\theta),
\end{align}
where $f(x)$ is the same as defined in Eq.~\eqref{eq:fr_gr}, and
\begin{align}
    \omega_2(\theta)=\frac{1}{32}-\frac{4f(2)-1}{64}\sin^2\theta.
\end{align}
Finally, the near and far expansions are
\begin{align}
    \psi_\mathrm{near}^{(2)}=\psi_\mathrm{mid}^{(2)}\big|_{x=2},&\quad \psi_\mathrm{far}^{(2)}=\psi_\mathrm{mid}^{(2)}\big|_{x\rightarrow\infty}, \\
    \chi^3 I_\mathrm{near}^{(5)}=\chi I_\mathrm{mid}^{(3)}\big|_{x=2},&\quad I_\mathrm{far}^{(2)}=\chi I_\mathrm{mid}^{(3)}\big|_{x\rightarrow\infty}, \\
    \chi^3 \Omega_\mathrm{near}^{(5)}=\chi \Omega_\mathrm{mid}^{(3)}\big|_{x=2},&\quad \Omega_\mathrm{far}^{(2)}=\chi \Omega_\mathrm{mid}^{(3)}\big|_{x\rightarrow\infty}. \label{eq:omega_near_far_gr}
\end{align}
Note that the first relative order solution vanishes, which supports the argument that the field variables should be smooth functions of $\chi$.
From Eqs.~\eqref{eq:psi0_mid_gr}--\eqref{eq:omega_near_far_gr}, it is clear that the near solutions are nothing but the mid solutions when taking $x=2$, as expected. Similarly, the far solutions are nothing but the mid solutions when taking $x\rightarrow\infty$. Therefore, the near and far expansions appear to be trivial up to the second relative order. In the following, we will solve the quadratic gravity corrections to the field variables, and we will show that the solutions have the same qualitative behavior as in GR.

\subsection{Leading Order in Spin}
Let us first consider the mid expansion. The stream Eq.~\eqref{eq:stream} reads
\begin{align}
    L\psi_{\mathrm{mid},q}^{(1,0)}=0,
\end{align}
where $L$ has been defined in Eq.~\eqref{eq:lop_mid}. We then require Eqs.~\eqref{eq:flux_pole} and \eqref{eq:flux_equator} as the boundary conditions in the angular direction. In the radial direction, matching the near and far expansions requires that $\psi^{(1,0)}_\mathrm{mid}$ be finite at both boundaries. 
The reason is the following: Suppose $\psi^{(1,0)}_\mathrm{mid}$ had some diverging dependence on $x$ as $x\rightarrow\infty$ which, for example, behaved like $x^{n}~(n>0)$. Then due to $x=\bar{x}/\chi$, there would have to be a corresponding $\psi^{(1,-n)}_\mathrm{far}$ in the far expansion. Given that $\psi=\mathcal{O}(1)$, there is no such $\psi^{(1,-n)}_\mathrm{far}$. Therefore, $\psi^{(1,0)}_\mathrm{mid}$ must be finite as $x\rightarrow\infty$. Similarly, one can also argue that $\psi^{(1,0)}_\mathrm{mid}$ must be finite as $x\rightarrow2$.
In the end, the solution has to be
\begin{align}
    \psi_{\mathrm{mid},q}^{(1,0)}=0.
\end{align}
The other two force-free conditions, Eqs.~\eqref{eq:ipsi} and \eqref{eq:omegapsi}, provide the following solutions:
\begin{align}
    I_{\mathrm{mid},q}^{(1,1)}=&\frac{\psi_0}{R_\mathrm{mid}}i_{0,q}(\theta),\\
    \Omega_{\mathrm{mid},q}^{(1,1)}=&\frac{1}{R_\mathrm{mid}}\omega_{0,q}(\theta),
\end{align}
where $i_{0,q}$ and $\omega_{0,q}$ are to be determined later. 

Next, we consider the near expansion. The stream Eq.~\eqref{eq:stream} reads
\begin{align}
    L_\mathrm{near}\psi_{\mathrm{near},q}^{(1,0)}=0,
\end{align}
where $L_\mathrm{near}$ is defined as~\cite{Armas:2020mio}
\begin{align}
    L_\mathrm{near}=16\partial_y+(-1+\cos2\theta+16y)\partial_y^2.
\end{align}
The angular boundary conditions are again Eqs.~\eqref{eq:flux_pole} and \eqref{eq:flux_equator}. On the horizon (i.e, $y=0$), the solution must follow Eq.~\eqref{eq:flux_horizon}. As $y\rightarrow\infty$, the solution must match the mid expansion; consequently, $\psi_{\mathrm{near},q}^{(1,0)}$ must be finite, and therefore
\begin{align}
    \psi_{\mathrm{near},q}^{(1,0)}=0.
\end{align}
Considering the other two force-free conditions, Eqs.~\eqref{eq:ipsi} and \eqref{eq:omegapsi}, together with the requirement that the solutions match the mid expansion, we obtain
\begin{align}
    I_{\mathrm{near},q}^{(1,3)}=&\frac{\psi_0}{R_\mathrm{near}}i_{0,q}(\theta),\\
    \Omega_{\mathrm{near},q}^{(1,3)}=&\frac{1}{R_\mathrm{near}}\omega_{0,q}(\theta).
\end{align}
We can now use the horizon Znajek condition and derive
\begin{align}
    i_{0,\mathrm{sGB}}(\theta)=&2\pi\left[\omega_{0,\mathrm{sGB}}(\theta)-\frac{21}{80}\right]\sin^2\theta, \label{eq:znajek0_near_gb}\\
    i_{0,\mathrm{dCS}}(\theta)=&2\pi\left[\omega_{0,\mathrm{dCS}}(\theta)+\frac{709}{28672}\right]\sin^2\theta. \label{eq:znajek0_near_cs}
\end{align}

Finally, we consider the far expansion. The stream equation [Eq.~\eqref{eq:stream}] reads:
\begin{align}
    &L_\mathrm{far}\psi_{\mathrm{far},q}^{(1,0)}-\frac{1}{32\sin\theta}\partial_\theta\left(\psi_{\mathrm{far},q}^{(1,0)}\cos\theta\right) \notag\\
    &=\frac{R_\mathrm{far}}{16\pi\sin\theta}\partial_\theta\left(I_{\mathrm{far},q}^{(1,0)}+2\pi\,\Omega_{\mathrm{far},q}^{(1,0)}\sin^2\theta\right), \label{eq:stream0_far_qg}
\end{align}
where $L_\mathrm{far}$ is defined as~\cite{Armas:2020mio}
\begin{align}
    L_\mathrm{far}=&\sin\theta\partial_\theta\left[\sin\theta\left(\frac{1}{\bar{x}^2\sin^2\theta}-\frac{1}{64}\right)\partial_\theta\right] \notag\\
    &+\sin^2\theta\partial_{\bar{x}}\left[\bar{x}^2\left(\frac{1}{\bar{x}^2\sin^2\theta}-\frac{1}{64}\right)\partial_{\bar{x}}\right] \notag\\
    &+\frac{1}{32}(2-3\sin^2\theta).
\end{align}
Because $\psi_\mathrm{far}$, $I_\mathrm{far}$, and $\Omega_\mathrm{far}$ are coupled, it is not easy to solve this equation directly. We propose the following ansatz: 
\begin{align}
    \psi_{\mathrm{far},q}^{(1,0)}&=0, \label{eq:psi0_far_qg_guess}\\
    I_{\mathrm{far},q}^{(1,0)}&=\frac{\psi_0}{R_\mathrm{far}} i_{0,q}(\theta), \label{eq:i0_far_qg_guess}\\
    \Omega_{\mathrm{far},q}^{(1,0)}&=\frac{1}{R_\mathrm{far}}\omega_{0,q}(\theta), \label{eq:omega0_far_qg_guess}
\end{align}
which satisfies the two force-free conditions Eqs.~\eqref{eq:ipsi}--\eqref{eq:omegapsi}, the boundary conditions Eqs.~\eqref{eq:flux_pole}--\eqref{eq:flux_equator} and \eqref{eq:flux_infinity}, and the condition that they match with the mid expansion.

We are now left with Eq.~\eqref{eq:stream0_far_qg} and the condition given by Eq.~\eqref{eq:znajek_infinity}. The latter requires
\begin{align}
    i_{0,q}(\theta)=-2\pi\,\omega_{0,q}(\theta)\sin^2\theta. \label{eq:znajek0_far_qg}
\end{align}
Inserting Eqs.~\eqref{eq:psi0_far_qg_guess}--\eqref{eq:znajek0_far_qg} into Eq.~\eqref{eq:stream0_far_qg}, we find that Eq.~\eqref{eq:stream0_far_qg} is also satisfied. Therefore, the proposed ansatz is indeed the solution. 

Now combining the conditions in Eqs.~\eqref{eq:znajek0_near_gb}--\eqref{eq:znajek0_near_cs} and \eqref{eq:znajek0_far_qg}, we determine $\omega_0$:
\begin{align}
    \omega_{0,\mathrm{sGB}}(\theta)=&\frac{21}{160}, \label{eq:omega0_sub_gb}\\
    \omega_{0,\mathrm{dCS}}(\theta)=&-\frac{709}{57344}. \label{eq:omega0_sub_cs}
\end{align}
Then, $i_0$ is given by Eq.~\eqref{eq:znajek0_far_qg}.

To summarize, at leading order in spin, we find
\begin{align}
    &\psi_{\mathrm{near},q}^{(1,0)}=\psi_{\mathrm{mid},q}^{(1,0)}=\psi_{\mathrm{far},q}^{(1,0)}=0, \\
    &\chi^3 I_{\mathrm{near},q}^{(1,3)}=\chi I_{\mathrm{mid},q}^{(1,1)}=I_{\mathrm{far},q}^{(1,0)}=-\frac{2\pi\psi_0 a}{M^2}\omega_{0,q}(\theta), \\
    &\chi^3\Omega_{\mathrm{near},q}^{(1,3)}=\chi\Omega_{\mathrm{mid},q}^{(1,1)}=\Omega_{\mathrm{far},q}^{(1,0)}=\frac{a}{M^2}\omega_{0,q}(\theta),
\end{align}
where $\omega_0$ is given in Eqs.~\eqref{eq:omega0_sub_gb} and \eqref{eq:omega0_sub_cs} in sGB and dCS gravity, respectively.

\subsection{First Relative Order in Spin}
We now go to next order. At first relative order, the mid-expansion stream equation [Eq.~\eqref{eq:stream}] reads
\begin{align}
    L\psi_{\mathrm{mid},q}^{(1,1)}=0.
\end{align}
We then require the boundary conditions in Eqs.~\eqref{eq:flux_pole}--\eqref{eq:flux_equator} and that they match with the other two expansions. The resulting solution is
\begin{align}
    \psi_{\mathrm{mid},q}^{(1,1)}=0,
\end{align}
while Eqs.~\eqref{eq:ipsi} and \eqref{eq:omegapsi} give
\begin{align}
    I_{\mathrm{mid},q}^{(1,2)}=&\frac{\psi_0}{R_\mathrm{mid}}i_{1,q}(\theta),\\
    \Omega_{\mathrm{mid},q}^{(1,2)}=&\frac{1}{R_\mathrm{mid}}\omega_{1,q}(\theta).
\end{align}

The near-expansion stream equation [Eq.~\eqref{eq:stream}] reads
\begin{align}
    L_\mathrm{near}\psi_{\mathrm{near},q}^{(1,1)}=0.
\end{align}
By requiring the boundary conditions in Eqs.~\eqref{eq:flux_pole}--\eqref{eq:flux_horizon} and that $\psi_\mathrm{near}$ match with $\psi_\mathrm{mid}$, we get
\begin{align}
    \psi_{\mathrm{near},q}^{(1,1)}=0, 
\end{align}
while Eqs.~\eqref{eq:ipsi} and~\eqref{eq:omegapsi} give
\begin{align}
    I_{\mathrm{near},q}^{(1,4)}=&\frac{\psi_0}{R_\mathrm{near}}i_{1,q}(\theta),\\
    \Omega_{\mathrm{near},q}^{(1,4)}=&\frac{1}{R_\mathrm{near}}\omega_{1,q}(\theta).
\end{align}
The condition in Eq.~\eqref{eq:znajek_horizon} can now be evaluated:
\begin{align}
    i_{1,q}(\theta)=&2\pi\,\omega_{1,q}(\theta)\sin^2\theta. \label{eq:znajek1_near}
\end{align}

The far expansion can be computed by starting from Eqs.~\eqref{eq:ipsi} and \eqref{eq:omegapsi}. The solutions are
\begin{align}
    I_{\mathrm{far},q}^{(1,1)}=&\frac{\psi_0}{R_\mathrm{far}}i_{1,q}-\frac{\pi}{2R_\mathrm{far}}\psi_{\mathrm{far},q}^{(1,1)}\cos\theta,\\
    \Omega_{\mathrm{far},q}^{(1,1)}=&\frac{1}{R_\mathrm{far}}\omega_{1,q}.
\end{align}
Then, the stream equation [Eq.~\eqref{eq:stream}] reads
\begin{align}
    &L_\mathrm{far}\psi_{\mathrm{far},q}^{(1,1)}=\frac{\psi_0}{16\pi\sin\theta}\partial_\theta\left(i_{1,q}+2\pi\sin^2\theta\,\omega_{1,q}\right). \label{eq:stream1_far_qg}
\end{align}
We propose the solution to be
\begin{align}
    \psi_{\mathrm{far},q}^{(1,1)}=0,
\end{align}
such that the condition in Eq.~\eqref{eq:znajek0_far_qg} becomes
\begin{align}
    i_{1,q}(\theta)=&-2\pi\,\omega_{1,q}(\theta)\sin^2\theta. \label{eq:znajek1_far}
\end{align}
Therefore, we can verify that Eq.~\eqref{eq:stream1_far_qg} is satisfied. 
Combining Eqs.~\eqref{eq:znajek1_near} and \eqref{eq:znajek1_far}, we have
\begin{align}
    i_{1,q}(\theta)=0=\omega_{1,q}(\theta).
\end{align}

To summarize, at first relative order we find
\begin{align}
    &\psi_{\mathrm{near},q}^{(1,1)}=\psi_{\mathrm{mid},q}^{(1,1)}=\psi_{\mathrm{far},q}^{(1,1)}=0, \label{eq:psi1_sum_qg} \\
    &I_{\mathrm{near},q}^{(1,4)}=I_{\mathrm{mid},q}^{(1,2)}=I_{\mathrm{far},q}^{(1,1)}=0, \\
    &\Omega_{\mathrm{near},q}^{(1,4)}=\Omega_{\mathrm{mid},q}^{(1,2)}=\Omega_{\mathrm{far},q}^{(1,1)}=0. \label{eq:omega1_sum_qg}
\end{align}

As these quadratic gravity corrections vanish, the field variables are still smooth functions of $\chi$ up to second relative order. 

\subsection{Second Relative Order in Spin}
At second relative order, the mid-expansion stream equation [Eq.~\eqref{eq:stream}] reads
\begin{align}
    L\psi_{\mathrm{mid},q}^{(1,2)}=\psi_0 s_q(x)\cos\theta\sin^2\theta.
\end{align}
Considering the boundary conditions in Eqs.~\eqref{eq:flux_pole}--\eqref{eq:flux_equator} and the matches with the other two expansions, the result takes the form
\begin{align}
    \psi_{\mathrm{mid},q}^{(1,2)}=\psi_0 h_q(x)\cos\theta\sin^2\theta,
\end{align}
where $h_q(x)$ is the solution to the radial equation
\begin{align}
    \frac{d}{dx}\left[\left(1-\frac{2}{x}\right)\frac{dh_q(x)}{dx}\right] - \frac{6h_q(x)}{x^2}=s_q(x),
\end{align}
with the boundary conditions such that $h_q(x)$ is finite at $x=2$ and when $x\rightarrow\infty$. The results are
\begin{widetext}
\begin{align}
    s_\mathrm{sGB}(x)
    =&-\frac{3}{4x}\left(1+\frac{1}{x}-\frac{44}{3 x^2}+\frac{34}{x^3}+\frac{16}{5 x^4}+\frac{976}{3 x^5}-\frac{448}{x^6}\right) \left[\mathrm{Li}_2\left(\frac{2}{x}\right)+\ln\left(\frac{2}{x}\right)\ln\left(1-\frac{2}{x}\right)\right] \notag\\
    &-\frac{3}{2(x-2)^2}\left(1-\frac{2}{x}-\frac{49}{3 x^2}+\frac{80}{x^3}-\frac{5296}{45 x^4}+\frac{15808}{45 x^5}-\frac{7092}{5 x^6}+\frac{80096}{45 x^7}-\frac{3424}{9 x^8}\right) \ln\left(\frac{2}{x}\right) \notag\\
    &+\frac{3}{2x(x-2)}\bigg(1-\frac{6}{5 x}-\frac{301}{18 x^2} + \frac{484756}{7875 x^3}-\frac{764041}{23625 x^4}+ \frac{50442368}{165375 x^5} -\frac{44345362}{55125 x^6}\notag\\ 
    &+ \frac{1993576}{33075 x^7}+ \frac{688420}{1323 x^8}-\frac{70712}{35 x^9}+ \frac{13856}{3 x^{10}}-\frac{3520}{x^{11}}\bigg), \label{eq:radial_gb}
\end{align}

\begin{align}
    h_\mathrm{sGB}(x)=&-\frac{8389 x^2}{60}+\frac{9649 x}{60}+\frac{74099}{2160}+\frac{12017}{720 x}-\frac{5331127}{1008000 x^2}-\frac{541351}{75600 x^3}-\frac{2652689}{176400 x^4}+\frac{125249}{36750 x^5}+\frac{451}{270 x^6}\notag\\
    &-\frac{73}{441 x^7}+\frac{32}{5 x^8}-\frac{40}{3 x^9}+\frac{1}{2520 (x-2) }\bigg(352338 x^3-986685x^2+488285x+129416+\frac{52143}{x} +\frac{1036}{x^2}\notag\\
    &+\frac{10438}{x^3}-\frac{69804}{x^4} +\frac{10272}{x^5}\bigg)\ln\left(\frac{2}{x}\right)+\frac{1}{240}\bigg(16778 x^3-35247 x^2+10110x+3120+\frac{1020}{x}+\frac{474}{x^2}\notag\\
    &+\frac{1168}{x^3} -\frac{1680}{x^4}\bigg)\bigg[-\mathrm{Li}_2\left(1-\frac{2}{x}\right)+\frac{\pi^2}{6}\bigg] -7(6x^2-3 x-1) \bigg[\mathrm{Li}_2\left(\frac{2}{x}\right) \ln\left(\frac{2}{x}\right)-2\mathrm{Li}_3\left(\frac{2}{x}\right)+2Z(3)\bigg] \notag\\
    &+21x^2(2x-3)\Bigg\{
    \frac{\pi^4}{90}+\frac{\pi^2}{12}\ln\left(1-\frac{2}{x}\right)\left[\ln\left(1-\frac{2}{x}\right)-2\ln \left(\frac{2}{x}\right)\right]+\frac{1}{24}\ln^2\left(1-\frac{2}{x}\right)\bigg[6\ln^2\left(\frac{2}{x}\right)\notag\\
    &+\ln^2\left(1-\frac{2}{x}\right)-4\ln\left(\frac{2}{x}\right)\ln\left(1-\frac{2}{x}\right)\bigg]
    +\frac{1}{4}\left[\mathrm{Li}_2\left(\frac{2}{x}\right)+\ln \left(\frac{2}{x}\right)\ln \left(1-\frac{2}{x}\right)\right]^2 \notag\\
    &+\ln\left(\frac{2}{x}\right)\bigg[\mathrm{Li}_3\left(1-\frac{2}{x}\right)-2Z(3)\bigg]+\bigg[\mathrm{Li}_4\left(\frac{2}{x}\right)-\mathrm{Li}_4\left(1-\frac{2}{x}\right)+\mathrm{Li}_4\left(\frac{2}{2-x}\right)\bigg]\Bigg\},
\end{align}

\begin{align}
    s_\mathrm{dCS}(x)
    =&\frac{709}{7168 x^3}+\frac{709}{3584 x^4}-\frac{71}{256 x^5}-\frac{303}{448 x^6}-\frac{3301}{3136 x^7} + \frac{1539}{112 x^8}-\frac{32763}{1568 x^9}-\frac{10341}{224 x^{10}}-\frac{270}{x^{11}},\\
    h_\mathrm{dCS}(x)
    =&\frac{709
    x^2}{14336}-\frac{709 x}{14336}-\frac{7799}{516096}-\frac{709}{21504
    x}+\frac{221699}{4300800 x^2}+\frac{147149}{1612800 x^3}+\frac{2261}{15360
    x^4}+\frac{7857}{31360 x^5}\notag\\&+\frac{1557}{1792 x^6}+\frac{3921}{3136 x^7}+\frac{27}{16 x^8}+\ln\left(\frac{2}{x}\right)\left(-\frac{709 x^2}{14336}+\frac{709x}{28672}+\frac{709}{86016}\right)\notag\\&+\bigg[\text{Li}_2\left(\frac{2}{x}\right)+\ln\left(\frac{2}{x}\right) \ln\left(\frac{x-2}{x}\right)\bigg]\left(-\frac{709 x^3}{28672}+\frac{2127 x^2}{57344}\right), \label{eq:radial_sol_cs}
\end{align}
\end{widetext}
where $\mathrm{Li}_n(x)\equiv\sum_{k=1}^\infty x^k/k^n$ is the polylogarithm function of order $n$, and $Z(x)\equiv\sum_{k=1}^\infty 1/k^{x}$ is the Riemann zeta function. 
At the boundaries, 
\begin{align}
    h_\mathrm{sGB}(2)=&-\frac{1865759261}{9408000}+\frac{11497\pi^2}{960}+\frac{49\pi^4}{60},\\
    h_\mathrm{dCS}(2)=&\frac{5562399}{40140800}-\frac{709\pi^2}{86016},
\end{align}
and
\begin{align}
    h_\mathrm{sGB}(x)\Big|_{x\rightarrow\infty}\sim&\frac{21}{80x}, \\
    h_\mathrm{dCS}(x)\Big|_{x\rightarrow\infty}\sim&-\frac{709}{28672x}.
\end{align}
Having $\psi_\mathrm{mid}^{(1,2)}$ solved, Eqs.~\eqref{eq:ipsi} and \eqref{eq:omegapsi} then give
\begin{align}
    I_{\mathrm{mid},q}^{(1,3)}=&\frac{\psi_0}{R_\mathrm{mid}}\Big[i_{2,q}(\theta) \notag\\
    &-4\pi\omega_0h_q(x)\sin^2\theta\cos^2\theta \notag\\
    &-4\pi\omega_{0,q}f(x)\sin^2\theta\cos^2\theta\Big],\\
    \Omega_{\mathrm{mid},q}^{(1,3)}=&\frac{1}{R_\mathrm{mid}}\omega_{2,q}(\theta),
\end{align}
where $\omega_0=1/8$ as given in Eq.~\eqref{eq:omega0_sub_gr}, and $\omega_0$ has been solved in Eqs.~\eqref{eq:omega0_sub_gb}--\eqref{eq:omega0_sub_cs}.

The near-expansion stream equation [Eq.~\eqref{eq:stream}] reads
\begin{align}
    L_\mathrm{near}\psi_{\mathrm{near},q}^{(1,2)}=0.
\end{align}
Requiring as boundary conditions Eqs.~\eqref{eq:flux_pole}--\eqref{eq:flux_horizon} and the match with the mid expansion, we get
\begin{align}
    \psi_{\mathrm{near},q}^{(1,2)}=\psi_0 h_q(2)\cos\theta\sin^2\theta,
\end{align}
while Eqs.~\eqref{eq:ipsi} and \eqref{eq:omegapsi} give
\begin{align}
    I_\mathrm{near,GB}^{(1,5)}=&\frac{\psi_0}{R_\mathrm{near}}\Big[i_{2,q}(\theta) \notag\\
    &-4\pi\omega_0h_q(2)\sin^2\theta\cos^2\theta \notag\\
    &-4\pi\omega_{0,q}f(2)\sin^2\theta\cos^2\theta\Big],\\
    \Omega_{\mathrm{near},q}^{(1,5)}=&\frac{1}{R_\mathrm{near}}\omega_{2,q}(\theta).
\end{align}
The horizon Znajek condition can now be evaluated:
\begin{align}
    i_{2,\mathrm{sGB}}=&2\pi\bigg\{\omega_{2,\mathrm{sGB}}+\frac{21103}{201600}\notag\\
    &+\omega_0\left[h_\mathrm{sGB}(2)-\frac{49}{128}\right]\sin^2\theta\notag\\
    &+\omega_{0,\mathrm{sGB}}\Big[f(2)-\frac{1}{4}\Big]\sin^2\theta\bigg\}\sin^2\theta. \label{eq:znajek2_near_gb}\\
    i_{2,\mathrm{dCS}}=&2\pi\bigg\{\omega_{2,\mathrm{dCS}}+\frac{169}{24576}\notag\\
    &+\omega_0h_\mathrm{dCS}(2)\sin^2\theta\notag\\
    &+\omega_{0,\mathrm{dCS}}\Big[f(2)-\frac{1}{4}\Big]\sin^2\theta\bigg\}\sin^2\theta. \label{eq:znajek2_near_cs}
\end{align}

In the far expansion, solutions to Eqs.~\eqref{eq:ipsi} and \eqref{eq:omegapsi} are
\begin{align}
    I_{\mathrm{far},q}^{(1,2)}=&\frac{\psi_0}{R_\mathrm{far}}i_{2,q}-\frac{\pi}{2R_\mathrm{far}}\psi_{\mathrm{far},q}^{(1,2)}\cos\theta,\\
    \Omega_{\mathrm{far},q}^{(1,2)}=&\frac{1}{R_\mathrm{far}}\omega_{2,q}.
\end{align}
Then, the stream Eq.~\eqref{eq:stream} reads
\begin{align}
    &L_\mathrm{far}\psi_{\mathrm{far},q}^{(1,2)}=\frac{\psi_0}{16\pi\sin\theta}\partial_\theta\left(i_{2,q}+2\pi\,\omega_{2,q}\sin^2\theta\right). \label{eq:stream2_far_qg}
\end{align}
We may guess that the solution is
\begin{align}
    \psi_{\mathrm{far},q}^{(1,2)}=0.
\end{align}
This way the condition~\eqref{eq:znajek0_far_qg} becomes
\begin{align}
    i_{2,q}=&-2\pi\,\omega_{2,q}\sin^2\theta. \label{eq:znajek2_far}
\end{align}
Therefore, we can verify that Eq.~\eqref{eq:stream2_far_qg} is satisfied. 
Combining Eqs.~\eqref{eq:znajek2_near_gb}--\eqref{eq:znajek2_near_cs} and \eqref{eq:znajek2_far}, we have
\begin{align}
    \omega_{2,\mathrm{sGB}}=&-\frac{21103}{403200}-\frac{1}{2}\omega_0\left[h_\mathrm{sGB}(2)-\frac{49}{128}\right]\sin^2\theta \notag\\
    &-\frac{1}{2}\omega_{0,\mathrm{sGB}}\Big[f(2)-\frac{1}{4}\Big]\sin^2\theta, \label{eq:omega2_sub_gb} \\
    \omega_{2,\mathrm{dCS}}=&-\frac{169}{49152}-\frac{1}{2}\omega_0h_\mathrm{dCS}(2)\sin^2\theta \notag\\
    &-\frac{1}{2}\omega_{0,\mathrm{dCS}}\Big[f(2)-\frac{1}{4}\Big]\sin^2\theta,\label{eq:omega2_sub_cs}
\end{align}
Then, $i_2$ is given by Eq.~\eqref{eq:znajek2_far}.

To summarize, at second relative order, we have
\begin{align}
    \psi_{\mathrm{mid},q}^{(1,2)}=&\psi_0 h_q(x)\cos\theta\sin^2\theta,\\
    I_{\mathrm{mid},q}^{(1,3)}=&\frac{\psi_0}{M}\bigg[i_{2,q}(\theta) \notag\\
    &-4\pi\omega_0h_q(x)\sin^2\theta\cos^2\theta\notag\\
    &-4\pi\omega_{0,q}f(x)\sin^2\theta\cos^2\theta\bigg], \\
    \Omega_{\mathrm{mid},q}^{(1,3)}=&\frac{1}{M}\omega_{2,q}(\theta),
\end{align}
in the mid expansion, where $\omega_2$ is given in Eqs.~\eqref{eq:omega2_sub_gb}--\eqref{eq:omega2_sub_cs}, and $i_2$ is related to $\omega_2$ by Eq.~\eqref{eq:znajek2_far}.

The solutions in the near and far expansions are just
\begin{align}
    \psi_{\mathrm{near},q}^{(1,2)}=\psi_{\mathrm{mid},q}^{(1,2)}\big|_{x=2},
    &\quad \psi_{\mathrm{far},q}^{(1,2)}=\psi_{\mathrm{mid},q}^{(1,2)}\big|_{x\rightarrow\infty}, \\
    \chi^2 I_{\mathrm{near},q}^{(1,5)}=I_{\mathrm{mid},q}^{(1,3)}\big|_{x=2},
    &\quad I_{\mathrm{far},q}^{(1,2)}=\chi I_{\mathrm{mid},q}^{(1,3)}\big|_{x\rightarrow\infty}, \\
    \chi^2 \Omega_{\mathrm{near},q}^{(1,5)}= \Omega_{\mathrm{mid},q}^{(1,3)}\big|_{x=2},
    &\quad \Omega_{\mathrm{far},q}^{(1,2)}=\chi \Omega_{\mathrm{mid},q}^{(1,3)}\big|_{x\rightarrow\infty}. \label{eq:omega_near_far_qg}
\end{align}
Therefore, the near and far solutions are nothing but the mid solutions when taking $x=2$ and $x\rightarrow\infty$, respectively. Therefore, the near and far expansions are still trivial up to  second relative order, allowing us to use the simpler method described in the main text to second relative order. The solutions presented in this appendix for the mid expansion coincide with the solutions presented in Sec.~\ref{sec:qg_bz0}--\ref{sec:qg_bz2}.


\bibliography{references}

\end{document}